\documentclass[aps,prl,reprint,superscriptaddress]{revtex4-1}
\usepackage{graphicx}
\usepackage{amsmath}
\usepackage{amsfonts}
\usepackage{bm}
\usepackage{hyperref}
\usepackage[usenames,dvipsnames]{color}
\usepackage[sf,big,noindentafter]{titlesec}
\titlespacing{\section}{0pt}{2ex}{1ex}
\titlespacing{\subsection}{0pt}{1ex}{0ex}
\titlespacing{\subsubsection}{0pt}{0.5ex}{0ex}

\newcommand{\beginsupplement}{%
        \setcounter{table}{0}
        \renewcommand{\thetable}{S\arabic{table}}%
        \setcounter{figure}{0}
        \renewcommand{\thefigure}{S\arabic{figure}}%
     }
     
\usepackage[table]{xcolor} 

\begin{document}
\title{Tissue flow induces cell shape changes during organogenesis}
\author{Gonca Erdemci-Tandogan}
\email{gerdemci@syr.edu}
\affiliation{Department of Physics, Syracuse University, Syracuse, New York 13244, USA}
\author{Madeline J. Clark}
\affiliation{Department of Cell and Developmental Biology, State University of New York, Upstate Medical University, Syracuse, New York 13210, USA}
\author{Jeffrey D. Amack}
\affiliation{Department of Cell and Developmental Biology, State University of New York, Upstate Medical University, Syracuse, New York 13210, USA}
\author{M. Lisa Manning}
\affiliation{Department of Physics, Syracuse University, Syracuse, New York 13244, USA}

\begin{abstract}
In embryonic development, cell shape changes are essential for building functional organs, but in many cases the mechanisms that precisely regulate these changes remain unknown. We propose that fluid-like drag forces generated by the motion of an organ through surrounding tissue could generate changes to its structure that are important for its function. To test this hypothesis, we study the zebrafish left-right organizer, Kupffer's vesicle (KV), using experiments and mathematical modeling. During development, monociliated cells that comprise the KV undergo region-specific shape changes along the anterior-posterior axis that are critical for KV function: anterior cells become long and thin, while posterior cells become short and squat. Here, we develop a mathematical vertex-like model for cell shapes, which incorporates both tissue rheology and cell motility, and constrain the model parameters using previously published rheological data for the zebrafish tailbud [Serwane \textit{et al.}] as well as our own measurements of the KV speed. We find that drag forces due to dynamics of cells surrounding the KV could be sufficient to drive KV cell shape changes during KV development. More broadly, these results suggest that cell shape changes could be driven by dynamic forces not typically considered in models or experiments.
\end{abstract}

\maketitle

\section*{Introduction}
In biological processes ranging from wound healing to cancer metastasis to embryogenesis, cells must move past one another in a dense environment. Sometimes groups of cells migrate collectively, such as neural crest cells in vertebrate embryos \cite{Kuriyama2014}, lateral line primordia in zebrafish \cite{Haas2006} or border cells in the Drosophila egg chamber \cite{Cai2016b}. In other cases, cells across a tissue intercalate or invaginate in processes like convergent extension~\cite{Yin2009} or gastrulation~\cite{Leptin1995}.

These motions involve mechanical processes that span orders of magnitude in length and time. On shorter length- and timescales, the distribution and activity of cytoskeletal and adhesion molecules within a single cell specify forces that control cell shapes~\cite{Maitre2012,Salbreux2012}. These forces can be well-modeled by vertex~\cite{Fletcher2014,Farhadifar2007a,Hufnagel2007a,Staple2010,Bi2015} or cellular potts~\cite{Graner1992,Kabla2012,Szabo2010} models, which often assume that cells in a tissue are in mechanical equilibrium. At longer timescales and lengthscales the assumption of mechanical equilbrium breaks down, as cells exchange neighbors and the tissue as a whole behaves like a  liquid~\cite{Steinberg1963,Foty1994,Forgacs1998,Foty2005,Beysens2000,Brodland2002,Smutny2017}. These large-scale behaviors are often described in terms of continuum~\cite{Etournay2015a} or active particle-based models~\cite{Ranft2010}. 

Recent work has attempted to bridge the gap between these two scales by extracting parameters for large-scale continuum models, such as the shear modulus or local deformation rate, from individual cell shapes~\cite{Etournay2015a,Bi2015,Bi2016}. However, there is surprisingly little work that goes in the other direction: investigating how slow dynamics at the scale of tissues might effect smaller-scale structures and cell shapes.

One exception is recent work by Cai \textit{et al.} that highlights the importance of global forces generated by slowly evolving dynamics of environment~\cite{Cai2016b}. Specifically, this work proposed that the migrating border cell cluster within the nurse cells of Drosophila egg chamber mimics the behavior of a sphere moving through a viscous fluid, and they found that drag forces due to the microenvironment of the migrating border cells strongly influence the cluster size and speed. While the focus of this study was on the size and speed, the authors also reported that the large clusters tend to be more elongated, leaving open the possibility that drag forces may influence the shape of the cluster. Additional work confirms that nurse cells also impose an oppositional force on the migratory cluster ~\cite{Aranjuez2016}.

A natural extension of these studies is to ask whether dynamic mechanical forces, such as drag, that are best understood as emergent properties of a large number of cells in a tissue can help drive specific shape changes in cells and organs that are important for their biological function. 

Developing embryos provide an excellent platform for testing this hypothesis.  It is well established that during embryogenesis individual cells undergo shape changes to generate emergent macroscopic patterns that are essential for building functional organs~\cite{Ramsdell2005,Gregor2005,Gibb2010}. So far, these shape changes have been largely explained by morphogen gradients or geometric constraints~\cite{Rauzi2013}.

\begin{figure}[h]
\centering
\includegraphics[width=3in]{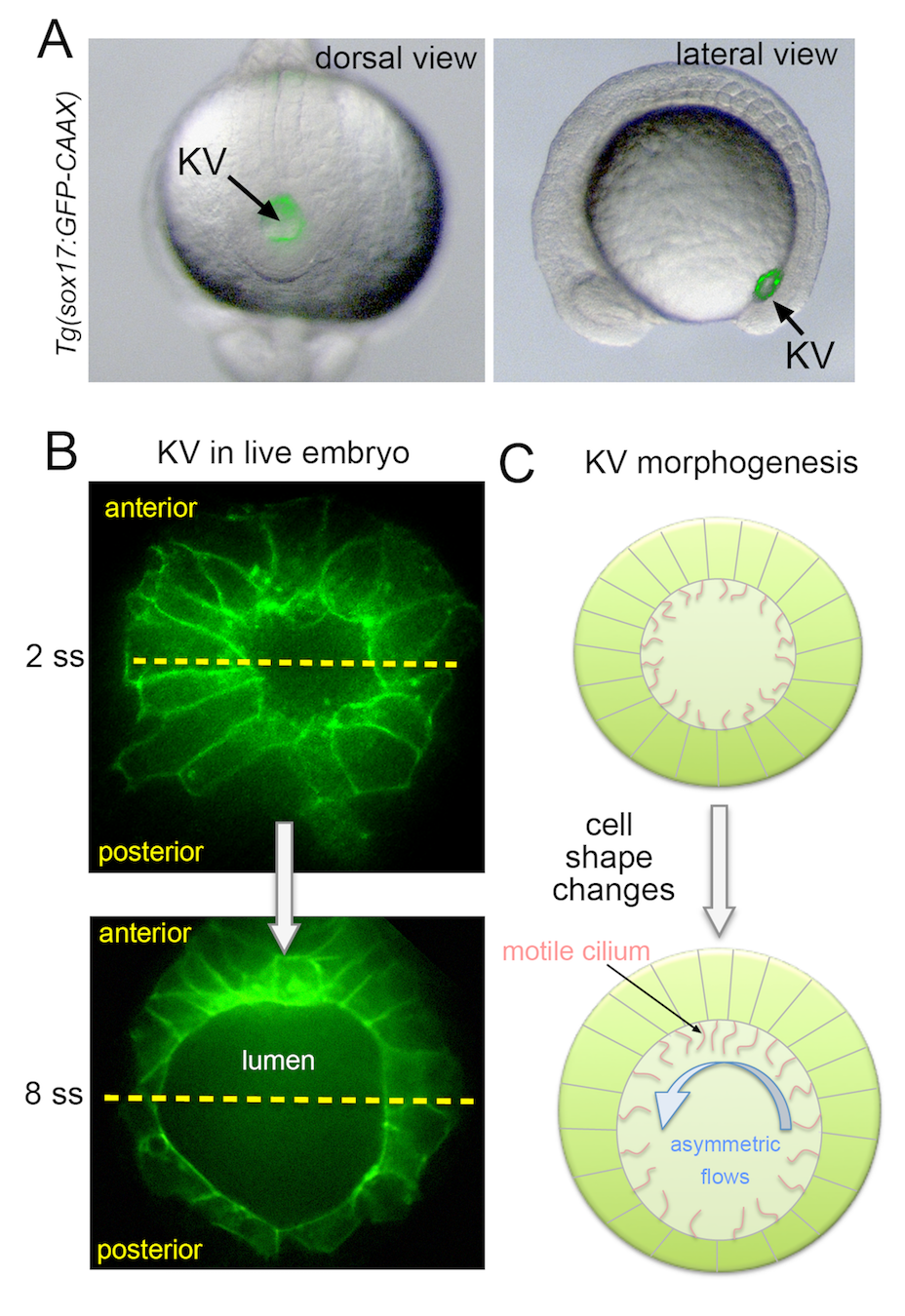}
\caption{\footnotesize{Cell shape changes during Kupffer's vesicle morphogenesis. (A) Expression of plasma membrane localized GFP (GFP-CAAX) labels KV (arrows) in transgenic Tg(sox17:GFP-CAAX) embryos. Dorsal and lateral views are shown. (B) Higher magnification dorsal views of labeled KV cells at 2 somite stage (ss) when all cells have similar morphologies and 8 ss when anterior cells are elongated and posterior cells are short and wide. Yellow lines bisect KV into anterior and posterior regions. (C) Schematic of how asymmetric cell shape changes along the anteroposterior axis concentrate motile cilia in the anterior region to generate asymmetric fluid flows that are critical for KV function in left-right patterning.}}
\label{KVremodeling} 
\end{figure}

In contrast, Kupffer's vesicle (KV) in the zebrafish embryo (Fig.~\ref{KVremodeling}A) provides an example of an organ that undergoes cell shape changes, and yet no morphogen gradient has yet been implicated. KV is a transient organ that helps to establish the left-right body plan \cite{Essner2005,Kramer-Zucker2005}. Motile cilia on KV cells  project into a fluid-filled lumen to generate asymmetric fluid flows that direct left-right patterning signals. We refer to KV and analogous ciliated structures in other vertebrates, which include the ventral node in mice and and gastrocoel roof plate in frogs, as left-right organizers \cite{Blum2014}. During morphogenesis of the KV organ, KV cells undergo region-specific shape changes that transform an initially symmetric structure to an organ with striking morphological differences along the anterior-posterior axis. Cells in the anterior region develop columnar morphologies whereas the cells located posteriorly become wide and thin (Fig.~\ref{KVremodeling}B-C). These cell shape changes result in more ciliated cells being tightly packed into the anterior region to promote asymmetric fluid flows in KV that define its function for the embryo (Fig.~\ref{KVremodeling}C)~\cite{Wang2011,Wang2012,Dasgupta2018}. 

One possibility is that these shape changes are driven by an unidentified morphogen gradient \cite{Dasbiswas2018}. Previous work~\cite{Wang2012} by some of us indicated that forces generated by myosin II within the KV help drive the shape change, and that perhaps there were different distributions of cytoskeletal molecules and forces across the anterior-posterior axis. Another possibility is highlighted by work from Compagnon \textit{et al.}~\cite{Compagnon2014a} suggesting that extra cellular matrix at the interface between the anteriorly-located notochord may also contribute to different mechanics on the anterior side of the organ. Additionally, recent work by some of us~\cite{Dasgupta2018} showed that ion flux mediates cell volume changes, which also contribute to the asymmetric cell shape changes observed in the KV.

While all these mechanisms could contribute to epithelial morphogenesis, here we identify another possibility that is suggested by studying time-sequence images of KV development. Specifically, the developing KV is roughly spherical, and it moves through the surrounding tissue as the embryo elongates along its anterior-posterior axis (Suppl. Movies 1 and 2). Migrating mesodermal tailbud cells are also highly dynamic, and very recent work has demonstrated that the tissue in the tailbud collectively behaves as a fluid~\cite{Lawton2013,Serwane2016a}. Moreover, the tailbud tissue near KV exhibits a strong retrograde flow near the KV~\cite{Lawton2013}. Taken together, these observations suggest that fluid-like drag forces generated by the motion of the KV through surrounding tissue may help drive changes to its structure.

To test whether such a mechanism could help drive the observed shape changes inside the KV, we develop a mathematical vertex-like model for the organ and surrounding tissue. We then constrain the model parameters using our \textit{in vivo} measurements of the KV speed through the surrounding tissue and previously published rheological data for the zebrafish tailbud \cite{Serwane2016a}.  We find that viscous flow of the cells surrounding the KV generates a drag similar to the one generated by a viscous fluid around a moving sphere.  These drag forces generate shape changes that qualitatively agree with the asymmetry observed in experiments: cells are columnar on the anterior side and cuboidal on the posterior side. Moreover, we find the forces generated by observed KV speeds are also of the correct order of magnitude to quantitatively match the data. More broadly, our results suggest that cell shape changes during embryonic development could be driven by large-scale dynamic forces not typically considered in models or experiments.

\section*{Methods}
\subsection*{Mathematical modeling of confluent tissue}
We study the KV motion through the surrounding tissue using a 2D vertex-like Self-Propelled Voronoi (SPV) model \cite{Bi2016}. Although Kupffer's vesicle is a fully three-dimensional organ, the mechanism we propose here is axisymmetric about the anterior-posterior axis, and so a simpler 2D model is sufficient to test the general behavior. Vertex or Voronoi models define an energy functional which describes cells in terms of their preferred geometry. For a confluent tissue containing N cells, this energy in 2D can be written as \cite{Fletcher2014,Farhadifar2007a,Hufnagel2007a,Staple2010,Bi2015}
\begin{equation}
E = \sum_{i}^{N} K_A (A_i-A_{0i})^2+K_P (P_i-P_{0i})^2+\sum_{{ij},{i<j}} \gamma_{ij} l_{ij}
\label{mechanicalenergy}
\end{equation}
The first term originates from cell volume incompressibility and monolayer's resistance to height fluctuations. $A_i$ and $A_{0i}$ are the actual and preferred cross-sectional areas of cell i. The term of the cell perimeter ($P_i$) is a result of subcortical actin cytoskeleton machinery and adhesive interactions between cells. These two competing forces balance to a preferred cell perimeter $P_{0i}$. $K_A$ and $K_P$ are the cell area and perimeter stiffness, respectively. The third term in the Eq.~\ref{mechanicalenergy} introduces an additional interfacial tension between different cell types $i$ and $j$. $\gamma_{ij}$ is the value of the interfacial tension and $l_{ij}$ is the length of the interface between adjacent cells $i$ and $j$ \cite{Sussman2018}.

To model the cell dynamics, we assign a constant self-propulsion speed $v_0$ along the direction of a cell polarization ${\bf \hat{n}_i}=(\cos \theta_i,\sin \theta_i)$. The dynamics of each cell is defined by the overdamped equation of motion of the cell centers $\mathit{\mathbf {r}}_i$
\begin{equation}
\frac{d \mathit{\mathbf {r}}_i}{d t}=\mu \mathbf {F}_{i} + v_0 {\bf \hat{n}_i}
\label{eqofmotion}
\end{equation}
where $\mathbf {F}_{i}=-\mathbf{\nabla}_i E$ is the mechanical force on cell $i$ and $\mu$ the mobility coefficient or the inverse microscopic friction. 

In principle, each of the cells in the simulation could be assigned a different self-propulsion speed to model individual cell motility. For simplicity, we choose to model the surrounding tailbud cells in the limit of vanishing motility ($v_0^{tailbud} = 0$), while the cells comprising the KV organ move collectively at the same, finite $v_0^{KV}$. Extending these methods to incorporate motility of tailbud cells shifts the fluidity of the surrounding tailbud cells in a well-characterized way~\cite{Bi2016}, and we do not expect it to significantly alter the results reported here.

The energy functional can be nondimensionalized by expressing all lengths in units of $\sqrt{A_0}$. This gives rise to a target shape index or a preferred perimeter-to-area ratio $p_0=P_0/\sqrt{A_0}$ which defines the tissue mechanical properties (material properties) exhibiting a transition from fluid-like tissue to solid-like tissue \cite{Bi2016}. While the observed cell area and perimeter values are available from the experimental measurements, we do not have direct access to the preferred cell shapes $p_0$. Therefore, we do not introduce any preferred asymmetries in the model to study solely the impact of cell dynamics on KV formation. Therefore, we set the preferred area ($A_{0i}=A_0$) and perimeter ($P_{0i}=P_0$) of all cells to be identical. There are two remaining independent parameters in the model: the self-propulsion speed $v_0$ and the cell shape index $p_0$.

To generate a cellular topology of KV and the surrounding tissue that is similar  observed in our experiments, we specify $2n$ KV cells (where $n$ is the number of anterior or posterior cells) which surround the lumen. The number of cells identified in 2D cross-sections of the KV midplane is about 15-20~\cite{Dasgupta2018}. Unless otherwise noted, we keep the total number of KV cells fixed at $2n=14$ for simulations reported here. The KV is embedded in $n_{ext}$ number of external or surrounding cells. Note that to ensure numerical stability in our simulations, we also represent the lumen as $n_{lumen}$ Voronoi cells. We keep the lumen area fixed for simplicity as previously shown that the cell shape changes do not depend on lumen expansion~\cite{Dasgupta2018}. We simulate this confluent tissue of total $N=2n+n_{ext}+n_{lumen}$ number of cells with periodic boundary conditions and using molecular dynamics \cite{Sussman2017b}. We conduct the simulations in two parts: First, we initialize the cellular structures for a static case in the absence of any cell motions and simulate the model using molecular dynamics with $10^5$ integration steps at step size $\triangle t=10^{-3}$ using Euler's method.  Second, we assign a self-propulsion speed to the KV cells $v_0=v_0^{KV}$ to mimic the dynamic motion of the KV through the tailbud cells and perform $10^5$ integration steps which is typically sufficient to drive the system to a steady state where the KV moves at constant velocity (see below). 

In the periodic box, we pin some random external cells ($<30$ cells) in order to avoid global flows of the entire tissue. Data presented in this study were computed from the average of 100 separate simulation runs. During integration steps, KV cells occasionally swap locations (T1 transitions within the KV cells) or lose contact with the lumen. These rearrangements are unphysical and do not exist in the experiments; thus, for averaging, we only consider KV cells without such events. We have also studied simulations where KV cells have a solid-like value of $p_0$ to prevent cell rearrangements entirely, and this does not change our conclusions. For simplicity, we focus here on simulations where all cells have the same $p_0$.

\subsection*{Measuring KV motion}
Transgenic Tg(sox17:GFP-CAAX)SNY101~\cite{Dasgupta2018} embryos were collected and incubated at 28.5$^{\circ}$C until the 1 somite stage (ss). Live embryos were positioned laterally in glass bottom dishes (MetTek) and immobilized in low melting point agarose. Immobilized embryos were left in their chorions to prevent the agarose from obstructing any cell movements. In a few cases, the embryo rotated inside the chorion during time-lapse imaging. These datasets were discarded. We found that most embryos did not move position during imaging, as shown in Suppl. Movie 1. Embryos were imaged using a Perkin-Elmer UltraVIEW Vox spinning disk confocal microscope while maintained at 32$^{\circ}$C. Images of the fluorescent KV (488 nm laser) and tailbud (differential interference contrast) were taken every five minutes at stages between 2 ss and 8 ss (a 3 hour window of development) using either a 10x objective or 20x objective. Image-J software (NIH) was used to create maximum projections and measure global speed of KV movement. 

To further visualize movement of KV and surrounding talibud cells, Tg(dusp6:memGFP)Pt19~\cite{Wang2011} or Tg(sox17:GFP-CAAX) embryos were injected with 50 pg mRNA  encoding mCherry with a nuclear localization sequence (NLS) at the one-cell stage. NLS-mCherry mRNA was synthesized in vitro from a linearized pCS2+ plasmid using Ambion mMessage mMachine SP6 Polymerase kit. Live embryos were immobilized in low melting point agarose and imaged using the spinning disk mircoscope at 32$^{\circ}$C for up to 2 hours between 2 ss and 8 ss. Images were captured every 2 minutes, and Image-J software (NIH) was used to create maximum projections. An example data set is shown in Suppl. Movie 2.
Experiments using zebrafish were approved by the Institutional Animal Care and Use Committee (IACUC) at SUNY Upstate Medical university. 

\begin{figure}[h]
\centering
\includegraphics[width=3in]{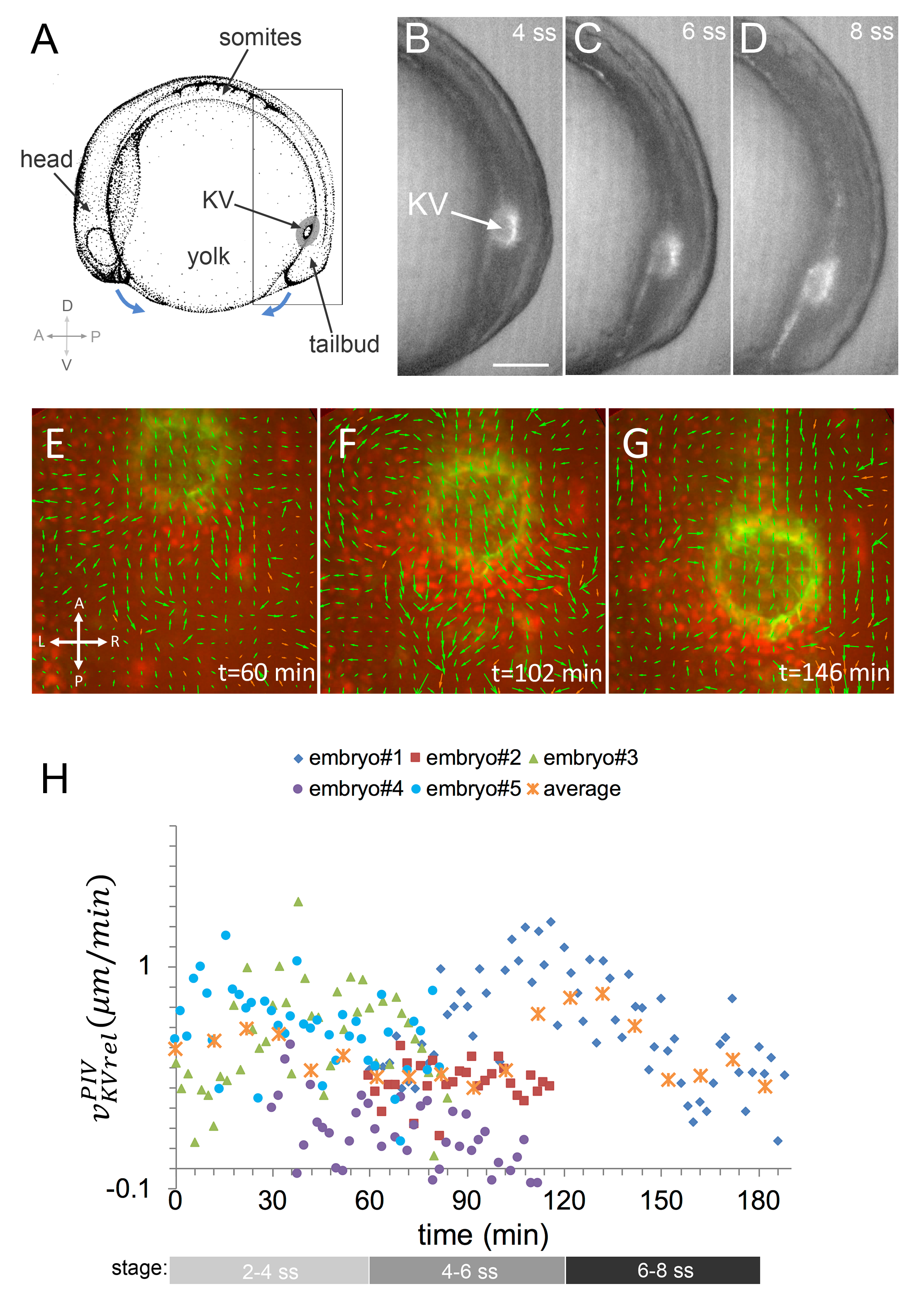}
\caption{\footnotesize{Motion of Kupffer's vesicle. (A) Schematic of a zebrafish embryo during early somite stages. Kupffer's vesicle (KV) develops in the tailbud of the embryo. Blue arrows represent body elongation along the anterior-posterior axis. D=dorsal, V=ventral, A=anterior, P=posterior. Boxed area corresponds to images in B-D. (B-D) Lateral view images of a live transgenic Tg(sox17:GFP-CAAX) embryo at 4 ss (B), 6 ss (C) and 8 ss (D) taken from a time lapse experiment. KV fluorescence (arrow) is superimposed on a DIC image of the tailbud region. (E-G) Snapshots from a time-lapse movie (Suppl. Movie 2) of a dorsal view of KV development from 4 ss to 8 ss. All nuclei are labeled red and KV cell membranes are labeled green. Green arrows are from PIV analysis indicating the direction of cell flow at grid points. (H) Scatter plot of KV velocity relative to tailbud along the posterior axis, taken at 2 min intervals from 5 embryos.  The orange x symbols denote the average of the data from all embryos binned into 10 min-intervals. t=0 corresponds to 2ss. Approximate somite stages are indicated in the gray bars. The images shown in (E-G) are maximum projections and do not highlight KV cell shapes in the mid-plane as shown in Fig.~\ref{KVremodeling}(B).}}
\label{KV} 
\end{figure}

\section*{Results}

\subsection*{Movement of Kupffer's vesicle during development}
The KV forms in the tailbud of the zebrafish embryo at the end of gastrulation and undergoes morphogenesis during early somite stages. At this time, the embryo is elongating along its anterior-posterior axis (Fig.~\ref{KV}A blue arrows). In order to measure the speed with which the KV moves during this process, we use two techniques. In one case, we perform a global analysis at relatively low resolution, where we used time-lapse imaging of transgenic Tg(sox17:GFP-CAAX)SNY101 embryos~\cite{Dasgupta2018} that mark KV cells with green fluorescence to visualize KV moving through the taildbud (Fig.~\ref{KV}B-D; Suppl. Movie 1). Although embryos were immobilized in agarose for imaging, they were left inside the chorion to allow normal cell movements. Measuring the speed of KV in five embryos indicated that the rate of KV movement remains similar between 2-8 somite stages (ss) (Suppl. Fig. 1 G; Suppl. Table 1). Since there were no statistically significant differences (Suppl. Table 1) in the KV speed as a function of time or between embryos, we average the speed for the entire range and for the five embryos. This results in an absolute average speed of the KV in the experiments $v_{KV}^{glob}= 1.47\pm0.71 \;\mu m/min$.

However, this is a measurement of the KV relative to a fixed reference frame such as the yolk. To determine the speed of KV relative to the tailbud tissue, which is the quantity that should govern the magnitude of drag forces experienced by the KV, we labeled all nuclei red by expressing nuclear-localized mCherry while labeling the KV cells by green fluorescent proteins in a wild-type Tg(dusp6:memGFP) embryo. Fig.~\ref{KV}(E-G) shows snapshots from a time-lapse movie (Suppl. Movie 2) of KV development from 4-8 somite stages of zebrafish development. Green arrows represent the direction of cell flow, calculated using Particle Image Velocimetry (PIV). The tailbud cells are highly mobile laterally and anteriorly. In contrast, the KV and notochord (which is adjacent to the KV on the anterior side) both move posteriorly.

To quantify the distinct behaviors of the KV and tailbud regions, we calculated the frame-to-frame displacements of the KV and the tailbud tissue using PIV (see Suppl. Methods for details). We measured the velocity of the KV (Suppl. Table 2) and the tailbud regions (Suppl. Table 3) in five embryos and found that the KV moves posteriorly more quickly than the surrounding tailbud tissue (Supp. Fig 1 A-E). We find that the absolute mean velocity of the KV along the posterior axis (Supp. Fig 1 F, Suppl. Table 2) is $v_{KV}^{PIV}= 1.04\pm 0.36 \;\mu m/min$, which is consistent with the global low-resolution measurement described above.  With the added resolution, we can also compute the mean velocity of tailbud cells, and most importantly, the relative velocity between the KV and tailbud regions along the posterior axis (see Suppl. Methods for details). Since there were no statistically significant differences from embryo to embryo (Fig.~\ref{KV}(H), Suppl. Table 4) or as a function of somite stage, we average this quantity over the entire temporal range for five embryos. We find the average velocity of KV relative to the tailbud along the posterior axis is $v_{KVrel}^{PIV}= 0.55\pm 0.30 \;\mu m/min$. As tailbud cells often have displacements perpendicular to the AP axis, we also quantified the mean relative velocity between the KV and tailbud regions (e.g. not restricted along the posterior axis), and found very similar results.

\subsection*{Constraining the simulation timescale}
We calibrate our modeling parameters using our experimental measurements of the KV speed relative to the tailbud (Fig.~\ref{KV} (H)) and previously published rheological data for the zebrafish tailbud tissue \cite{Serwane2016a}.

To compare our simulation time units with time in experiments, we first develop a vertex model that recapitulates the droplet deformation experiments of Serwane \textit{et al.} \cite{Serwane2016a} that they used to quantify relaxation timescales in zebrafish tailbud tissue~\textit{in vivo}. Specifically, they inject a single ferrofluid droplet between the cells of the progenitor zone (PZ) or presomitic mesoderm (PSM) of developing zebrafish embryos. When the droplet is actuated, it generates a local force dipole in the tissue deforming only the surrounding neighboring cells by a small amount without any apparent cellular rearrangements. By analyzing the temporal response of the tissue to this force, they identified two relaxation timescales: $1$ $sec$ and $1$ $min$.

\begin{figure}[h]
\centering
\includegraphics[width=3in]{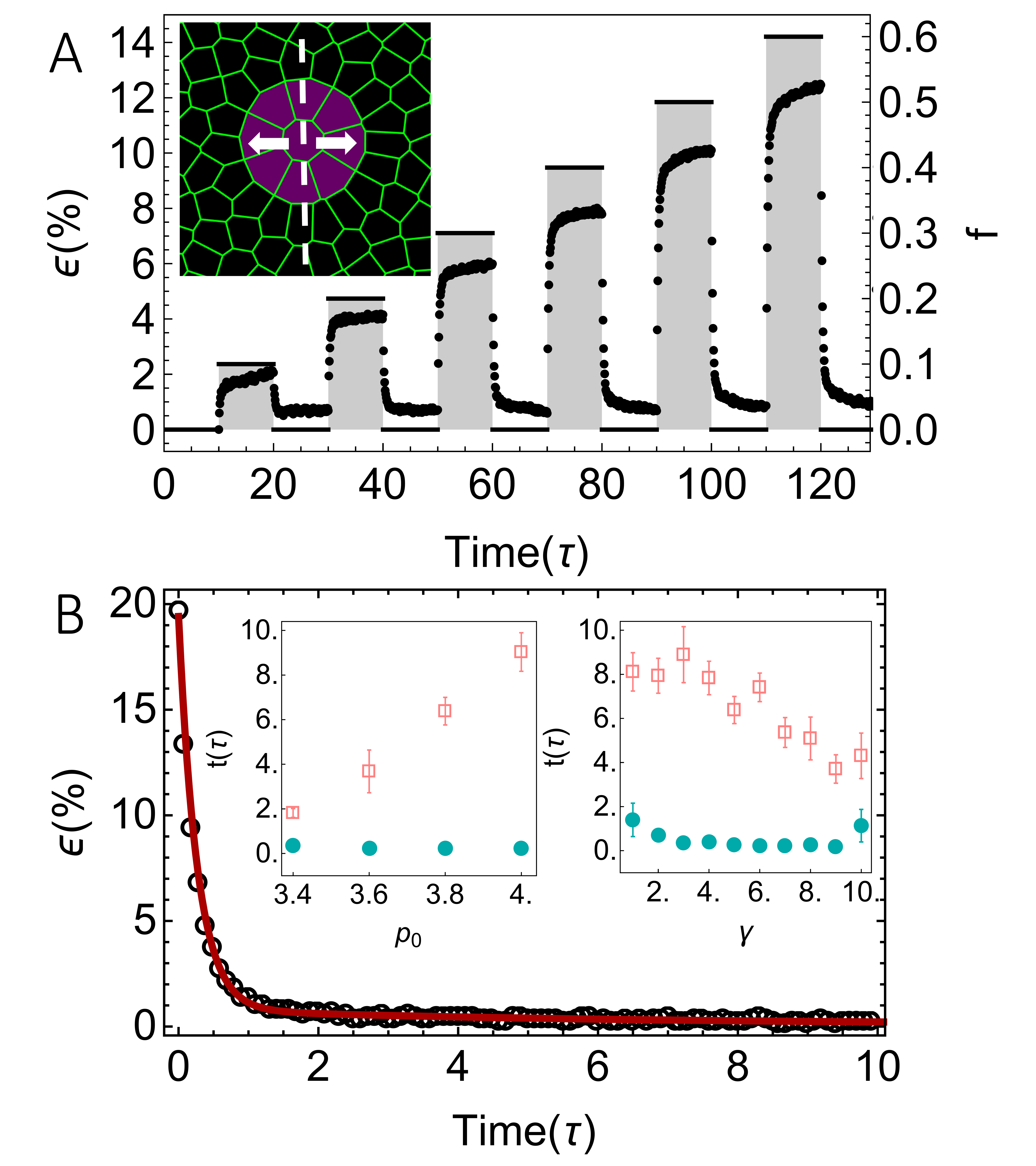}
\caption{\footnotesize{(A) Strain evolution over a series of actuation cycles. Shaded regions correspond to times where the constant force $f$ was applied. $f$ deforms the droplet until a force balanced state is reached (saturation of the strain). The droplet relaxes back to its initial shape after the force is turned off. Inset of (A) represents topology of the simulations. Droplet is represented as a supracell (purple colored cells) made of Voronoi units and embedded in the surrounded neighboring tissue (black colored cells). Dashed line denotes the mid-section of the droplet and arrows represents the direction of the deformation. (B) An example strain relaxation curve (circles) fits to a rheological model (line) with two characteristic relaxation timescales: $t_1=0.27\; \tau$ and $t_2=6.15\; \tau$ for $p_0=3.8$, $p_0^{droplet}=3.8$, $\gamma=5.0$ and $f=1.0$. Inset of (B) is the relaxation timescales (filled green circles represent shorter timescale $t_1$ while the open pink squares represent the longer timescale $t_2$) for various target shape index $p_0$ and droplet surface tension $\gamma$ values. Error bars that are not visible are less than $0.1$. Other simulation parameters are $K_A=100$, $K_P=1.0$, $\mu=1$, $p_0^{droplet}=3.8$, $A_0=1$, $n_{ext}=400$ (number of external cells, black colored cells of (A), not all number of cells are shown) and $n_{droplet}=9$ (number of droplet Voronoi units, purple colored cells).}}
\label{actuationcycles} 
\end{figure}

Using Eq. 1, we simulate a droplet embedded in a confluent tissue of 400 cells. Inset of Fig.~\ref{actuationcycles} (A) is an example initial configuration. We represent the droplet as several Voronoi cells to prevent numerical instabilities. The Voronoi-cell representation of the droplet does not affect the results presented here, as we keep the target shape index of the droplet Voronoi cells fixed in fluid-like regime $p_0^{droplet}=3.8$ where they do not contribute the shear modulus, as the effective interfacial tension between the cell interfaces in the fluid regime is zero~\cite{Yang2017}. We define an explicit line tension $\gamma$ as introduced in Eq. 1 to distinguish the droplet cells from the external cells. $\gamma$ can also be considered as the surface tension of the droplet. We then perform a rheology simulation by deforming the droplet from its original spherical shape. The droplet is deformed without a net force following the equation of motion
\begin{equation}
\frac{d \mathit{\mathbf {r}}_i}{d t}=\mu( \mathbf {F}_{i} + \mathbf {F}_{di})
\label{eqofmotiondroplet}
\end{equation}
where
\begin{equation}
 \mathbf{ F}_{di}=\begin{cases}
    f*(x_i-x_{cm})/\mu \;\;\hat{x}, & \text{for droplet cells}.\\
    0, & \text{otherwise}.
  \end{cases}
\label{fdroplet}
\end{equation}
and $\mathbf {F}_{i}=-\mathbf{\nabla}_i E$ is the gradient of the energy functional introduced in Eq.1. $f$ is the constant applied force that deforms the droplet along the direction $\hat{x}$ and $x_{cm}$ is the center of mass of the droplet. Upon application of $f$, the droplet relaxes to an ellipsoid along the direction of $f$ and likewise it goes back to its original spherical shape after turning off the force $f$ (Suppl. Movie 3). We characterize the deformation by a strain $\epsilon=(b-a)/a$ where $a$ and $b$ are the major and minor axis of the ellipsoid.
An example strain evolution over series of actuation cycles are shown in Fig.~\ref{actuationcycles} (A) where the shaded regions correspond to times where the constant force $f$ was applied.

The relaxation in strain as a function of time for a single cycle, shown in Fig.~\ref{actuationcycles} (B), displays at least two characteristic relaxation timescales. It is common to model such bimodal relaxation processes using systems of springs and dashpots, and here we are able to best fit the response using a generalized Maxwell fluid composed of two dashpot elements and a single spring. For a fixed $p_0=3.8$, $p_0^{droplet}=3.8$, $\gamma=5.0$ and $f=1.0$, we average over 100 separate simulation runs and find two characteristic relaxation timescales: $t_1=0.25\pm0.01\; \tau$ and $t_2=6.38\pm0.62\; \tau$ (average $\pm$ one standard error) where $\tau$ is the natural time unit of the simulation. Rarely, unphysical T1 transitions occurred between the artificial Voronoi cells within the droplet and generated spurious forces \cite{Sussman2018}. Hence, for averaging, we only take into account the droplets without such rearrangements.

The spring element of the rheological model is an effective spring composed of the elasticity of the droplet and the elasticity of the external tissue surrounding the droplet. Therefore, we next test whether the timescales we find depend on $p_0$ and $\gamma$ values. As illustrated in Fig.~\ref{actuationcycles} (B) inset, different $p_0$ and $\gamma$ parameters yield the same order of magnitude relaxation scales, although larger values of $p_0$ and smaller values of $\gamma$ are associated with longer relaxation times.

Thus, deforming a confluent tissue by a small amount we obtain a bimodal relaxation as $t_1=0.1\;\tau$ and $t_2=1\;\tau$ to within an order of magnitude. We compare the longer relaxation time scale of the simulations, $t_2$, with the longer timescale reported for the zebrafish tailbud tissue, $1$ $min$ \cite{Serwane2016a}. This gives us a rough approximation for the natural unit of the simulations as $\tau\sim1$ $min$.

As presented above, we found the average KV speed relative to the tailbud tissue as $v_{KVrel}^{PIV}= 0.55\pm 0.30 \;\mu m/min$ in the experiments. To convert this value to our simulation units, we also estimate a cell diameter as $10 \; \mu m$. Consequently the natural length scale of our simulations becomes $l=\sqrt{A_0}\sim10 \; \mu m$.   Taken all together, an order of magnitude estimate for the speed of KV in simulation units corresponds to $v_{KVrel}^{PIV}= 0.055 \pm 0.030 \; l/\tau$. We note the reported error bars only reflect the fluctuations in the experimental measurement, and do not reflect the additional uncertainties introduced by our conversion from experimental to simulation units.

\subsection*{Mathematical modeling indicates that dynamic motion of organ can drive cell shape change}
\begin{figure}[h]
\centering
   \includegraphics[width=3in]{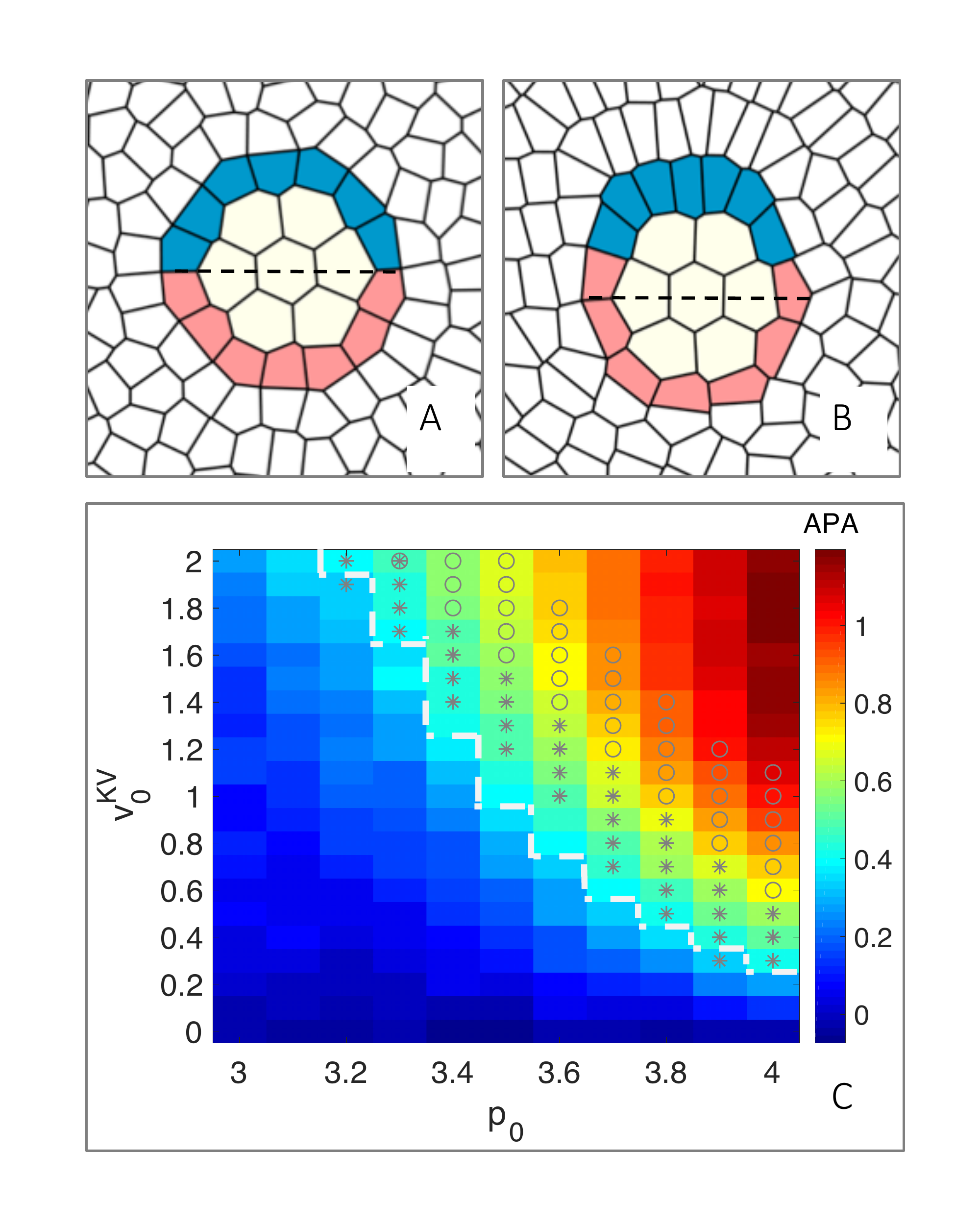}
  \caption{\footnotesize{ (A) KV remains symmetric in the absence of directed KV motion. (B) A snapshot from a simulation with directed KV motion demonstrates that the dynamics can generate asymmetric cell shape changes, and the asymmetry is similar to that seen in the experiments: posterior cells become short and squat, while those on the anterior side become elongated. The asymmetric cell shapes are accompanied with dense packing of the KV cells anteriorly; total number of cells positioned above the dashed line increases. Blue: anterior KV cells, pink: posterior KV cells, yellow: lumen Voronoi units, white: external cells. (C) Parameter scan for anterior posterior asymmetry, $APA=LWR_{ant}-LWR_{pos}$, as a function of self-propulsion speed of the KV cells ($v_0^{KV}$) and target shape index of all cells ($p_0$). The region to the upper right of the dashed white line corresponds to the range of experimental APA values of $0.91\pm0.51$. Asterisks correspond to the the average relative velocity of KV in the experiments in natural units ($v_{KVrel}^{PIV}= 0.055 \pm 0.030 \; l/\tau$) and gray circles correspond to the global measurement of mean velocity in natural units ($v_{KV}^{glob}= 0.147\pm0.071 \; l/\tau$). Number of anterior or posterior cells $n=7$, number of external cells $n_{ext}=400$, and number of Voronoi units $n_{lumen}=7$. Other simulation parameters are $K_A=100$, $K_P=1.0$, $\mu=1$, $A_0=1$, line tension between the KV cells and external cells $\gamma_{KV-ext}=1$, line tension between the KV cells and lumen cells, $\gamma_{KV-lumen}=1$ and line tension between the external cells and lumen cells, $\gamma_{ext-lumen}=100$.}}
  \label{APA} 
 \end{figure}

We next use our mathematical model to study the morphology of the KV as it moves through the tailbud tissue.  Unlike other proposed mechanisms for KV asymmetry, which all involve explicit, time-invariant mechanical asymmetries~\cite{Wang2012,Compagnon2014a,Dasgupta2018}, the KV model studied here contains no explicit mechanical asymmetry, and so the KV remains perfectly symmetric in the absence of directed KV motion. This is illustrated by the simulation snapshot in Fig~\ref{APA}(A).

However, when the KV is propelled through surrounding tissue, the tissue generates dynamic forces on the KV that are asymmetric -- the forces experienced at the ``front" of the moving object are different from those experienced at the ``rear".  A snapshot from a simulation with directed KV motion, Fig~\ref{APA}(B), demonstrates that directed motion of the KV can generate asymmetric cell shape changes, and the asymmetry is similar to that seen in experiments: posterior cells that ``at the front end" of the moving KV become short and squat, while those on the anterior side become elongated (Supplementry Movie 4). As we do not know the exact mechanisms of the KV movement (although possibilities are highlighted in the Discussion), we have also simulated cases where the KV is not directly self-propelled, but instead indirect forces generated by convergent extension of nearby tissue drive it through the surrounding tissue (Supplementry Materials). Our results suggest that movement of the KV creates asymmetric cell shape changes regardless of the mechanism that drives the KV through the tissue.

It is not clear if the parameters we have chosen for the model are relevant or reasonable for the \textit{in vivo} motion of the KV.  The benefit of mathematical modeling is that we can next study how the asymmetry depends on the material properties of the surrounding tailbud tissue, and whether the experimentally observed KV speed is the right order of magnitude to generate the shape changes. 

To analyze how the material properties of the tailbud tissue influence KV shape remodeling, we vary the target shape index $p_0$. Larger values of $p_0$ correspond to elongated, anisotropic cell shapes, while smaller values of $p_0$ correspond to isotropic, roundish shapes. Previous work has shown that $p_0$ controls the fluidity of the surrounding tissue; when the individual motility of tailbud cells is vanishingly small as assumed here, the tissue is solid like for $p_0 < 3.8$ and fluid-like for $p_0> 3.8$~\cite{Bi2016}.  

To compare simulations with experimental data quantitatively, we use a metric introduced previously \cite{Dasgupta2018}, anterior posterior asymmetry (APA). APA is the average difference between the length to width ratios (LWR) of anterior and posterior cells ($APA=LWR_{ant}-LWR_{pos}$). Previous work showed that APA changes from approximately $0$ to $0.91\pm0.51$ within one standard deviation of the average between 2 and 8 somite stages of zebrafish development \cite{Dasgupta2018} (from 12 embryos). 

Figure~\ref{APA} (C) is a plot of steady-state APA as a function of the self-propulsion speed of KV cells ($v_0^{KV}$) and target shape index of all cells ($p_0$). Our first observation is that there is a region, in the upper right hand side of the phase diagram, where the APA achieves values similar to the APA values observed \textit{in vivo} experiments (Fig.~\ref{APA} (C) the region to the upper right of the dashed white line). Clearly, higher APA values correlate with more fluidity of the tailbud cells (higher values of $p_0$) and larger values for the KV propulsion.

In our model, the parameter $v_0^{KV}$ specifies the rate at which the KV would move if it were not interacting with any of the tailbud cells.  However, interactions with the tailbud can slow the KV if the tailbud cells act like a fluid, or they can even stop the KV if the tailbud acts like a solid, as the elastic response can balance the propulsive force. Therefore, we explicitly calculate the speed of the KV in simulations for direct comparison to experimental measurements. The steady-state velocity, $v_{ss}$, values are presented in Supp. Fig. 3. The region marked by asterisks in Fig.~\ref{APA} (C) and in Supp. Fig. 3 corresponds to the average relative velocity of KV in the experiments in natural units ($v_{KVrel}^{PIV}= 0.055 \pm 0.030 \; l/\tau$). Since our measurements of KV velocity relative to the tailbud tissue averages over a rather large tailbud region and therefore likely underestimates the relative velocities, here we also mark our global low-resolution measurement of mean velocity in natural units $v_{KV}^{glob}= 0.147\pm0.071 \; l/\tau$ as an upper-limit estimate (Fig.~\ref{APA} (C) region marked by gray circles). There is a region of parameter space where both experimental APA values and $v_{ss}$ match the values seen in experiments, illustrated as boxes with asterisks and circles in Fig.~\ref{APA} (C).

An example simulation from this regime with $p_0=3.8$ is shown in Suppl. Movie 4, where the APA asymptotes to a value of $APA=0.96$ and an equilibrium speed of $v_{ss}=0.21$ after the KV cells are propelled through the surrounding tissue ($v_0^{KV}=1.4$).
\subsection*{Cell shape change is accompanied by asymmetric cilia distribution}
Optical cross-sections through the mid-plane of KV show that the LWRs become significantly different between anterior and posterior cells over development \cite{Wang2011,Wang2012,Dasgupta2018}. Furthermore, the cells are densely packed in the anterior half of KV resulting AP asymmetric cilia distribution. In other words, more cilia are positioned in the anterior region. This difference in the number of cilia between anterior and posterior poles has been shown to be required for asymmetric fluid flows in the lumen that are necessary for the proper left right patterning of the embryo \cite{Wang2012}.

 \begin{figure}
 \centering
   \includegraphics[width=3in]{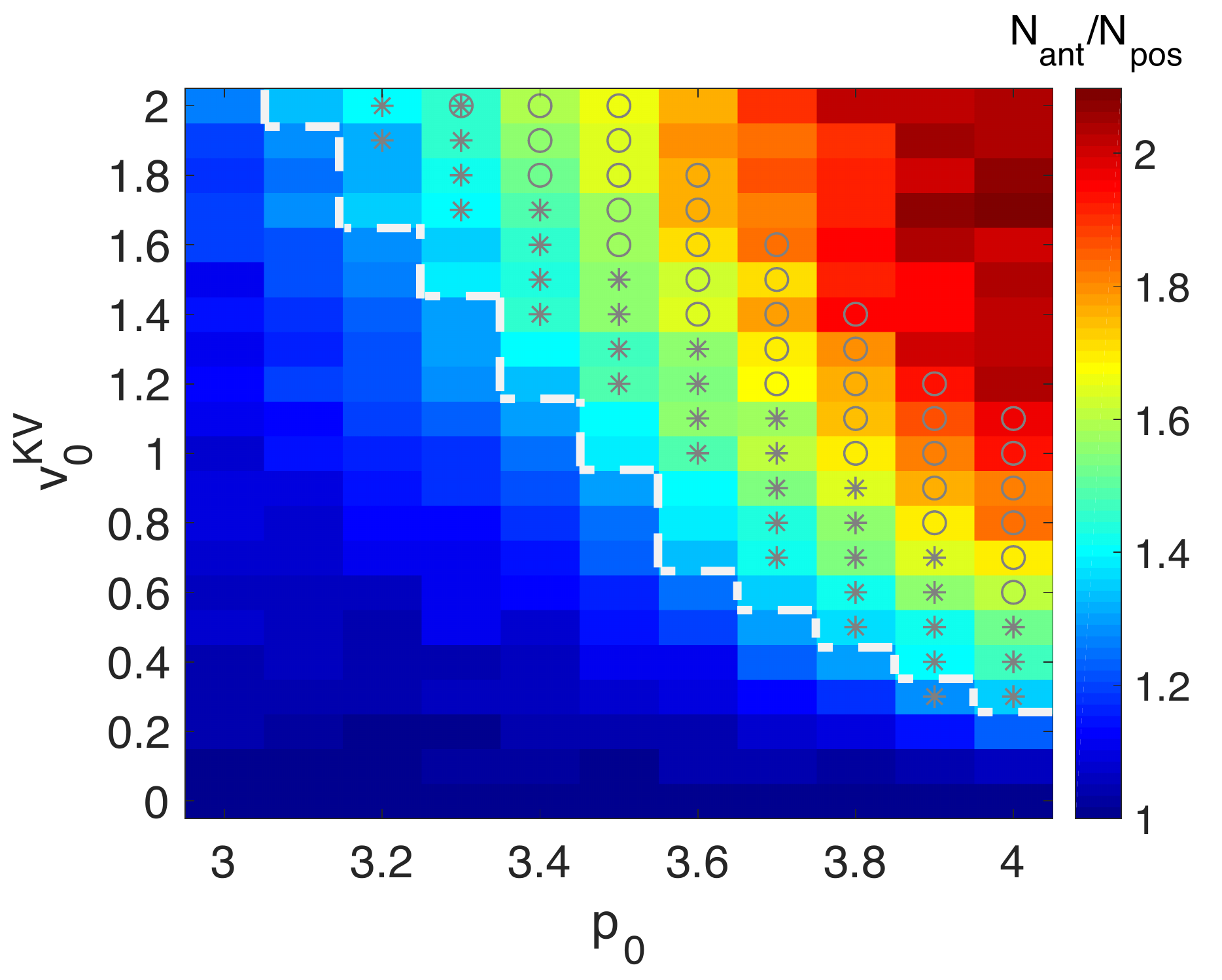}
  \caption{\footnotesize{Parameter scan for the ratio of number of cells above ($N_{ant}$) and below ($N_{pos}$) the KV mid-plane (Fig.~\ref{APA} (A) $\&$ (B) dashed lines) as a function of self-propulsion speed of the KV cells ($v_0^{KV}$) and target shape index of all cells ($p_0$). The region to the upper right hand of the white dashed line corresponds to the range of experimental $N_{ant}/N_{pos}$ values of $1.80\pm0.49$ at 8-10 ss. The region marked by asterisks corresponds to the average relative velocity of KV in the experiments in natural units ($v_{KVrel}^{PIV}= 0.055 \pm 0.030 \; l/\tau$/) and gray circles correspond to the global measurement of mean velocity in natural units ($v_{KV}^{glob}= 0.147\pm0.071 \; l/\tau$). Other simulation parameters are same as in Fig.~\ref{APA}.}}
  \label{cilia} 
 \end{figure}

In the absence of cell dynamics, mathematical modeling indicates that cells do not become as asymmetric as in wild-type experiments and they are not densely packed at one side of the organ (Fig.~\ref{APA} (A)), even though the cells can change shape slightly under certain conditions \cite{Dasgupta2018}. Interestingly, KV motion towards the tailbud cells could alone drive the asymmetries, resulting in both high APA values and the tight packing of more ciliated cells into the anterior region similar to the effects seen in wild-type experiments. As shown in Fig.~\ref{APA} (B), more KV cells position above the dashed line as a result of the surrounding tissue flow. 

We analyze this observation systematically for a parameter sweep of $v_0^{KV}$ and $p_0$. We count the number of cells above ($N_{ant}$) and below ($N_{pos}$) the mid-KV, which is the center of mass of the organ (Fig.~\ref{APA} (A $\&$ B) dashed line). Our \textit{in vivo} experiments indicate that ratio of $N_{ant}/N_{pos}=1.14\pm0.19$ at 3-4 ss (from 9 embryos) and it increases to $1.80\pm0.49$ at 8-10 ss (from 13 embryos) within one standard deviation of the ratio average. We note that we calculate the standard deviation directly from the ratio rather than from the variation of the number of anterior and posterior cells since previous studies found considerable variability in the number of KV cilia or cells \cite{Wang2011,Wang2012} in wild-type embryos but no effect on the cell shape changes. In simulations, the $N_{ant}/N_{pos}$ ratio takes values similar to the values observed \textit{in vivo} experiments (Fig.~\ref{cilia} the region to the upper right of the white dashed line). The region marked by asterisks corresponds to the average relative velocity of KV in the experiments in natural units ($v_{KVrel}^{PIV}= 0.055 \pm 0.030 \; l/\tau$) and the gray circles corresponds to the global measurement of mean velocity in natural units $v_{KV}^{glob}= 0.147\pm0.071 \; l/\tau$,  coinciding with a similar region of the parameter space presented in Fig.~\ref{APA} (C), where APA values of the simulations are similar to the APA values observed in the experiments.

We have repeated the simulations for larger KV, with more cells and a larger lumen (with up to $24$ KV cells), as the KV exhibits natural size variation in the experiments \cite{Roxo-Rosa2015}. Regardless of the size, the KV becomes highly asymmetric as it moves through the surrounding tissue, and the range of asymmetries are within the range of experimental observations.

\subsection*{Tissue drag forces govern cell shape change}
Although there are uncertainties in both the estimate of our simulation timescale $\tau$ and the KV speed measured in experiments, our model suggests that the experimentally observed speeds are the right order of magnitude to dynamically generate the cell shape asymmetries seen in the KV.

Next, we wonder whether there is a simple physical explanation for the mechanism driving this shape change. The observation in Fig.~\ref{APA} (C) that the simulations are closest to experiments in the upper-right hand corner where the environment is fluid-like, suggest an answer.  Specifically, we hypothesize that the viscous flow of tailbud cells around the roughly spherical KV generates a drag force that reshapes the KV, blunting the shape in the front (which in this case corresponds to the posterior side of the KV) and extending it in the rear (anterior side).

To test this hypothesis, we note that an object moving in a viscous fluid experiences a drag force in the direction opposite to its motion. The effective force on the moving KV sphere, in a fluid-like material, can be written as $F_{eff}=\sum_{i=1}^{2N} \frac{v_0^{KV}}{\mu}- b  v$ where the first term on the right stands for the self-propulsion of the KV cells and the second term is an emergent, frictional drag. Here $\mu$ is a model parameter that represents the inverse of the microscopic friction, $b$ the emergent, macroscopic friction and $v$ the observed speed of the KV in our simulations. The terminal velocity (steady-state speed) is achieved when the drag force is equal in magnitude to the force propelling the object which in this case self-propulsion force of the KV. Thus, one can relate the drag force in terms of model the parameters at balanced state: $F_{drag}=b  v_{ss}=\sum_{i=1}^{2N} \frac{v_0^{KV}}{\mu}$.

 \begin{figure}[h]
 \centering
   \includegraphics[width=3in]{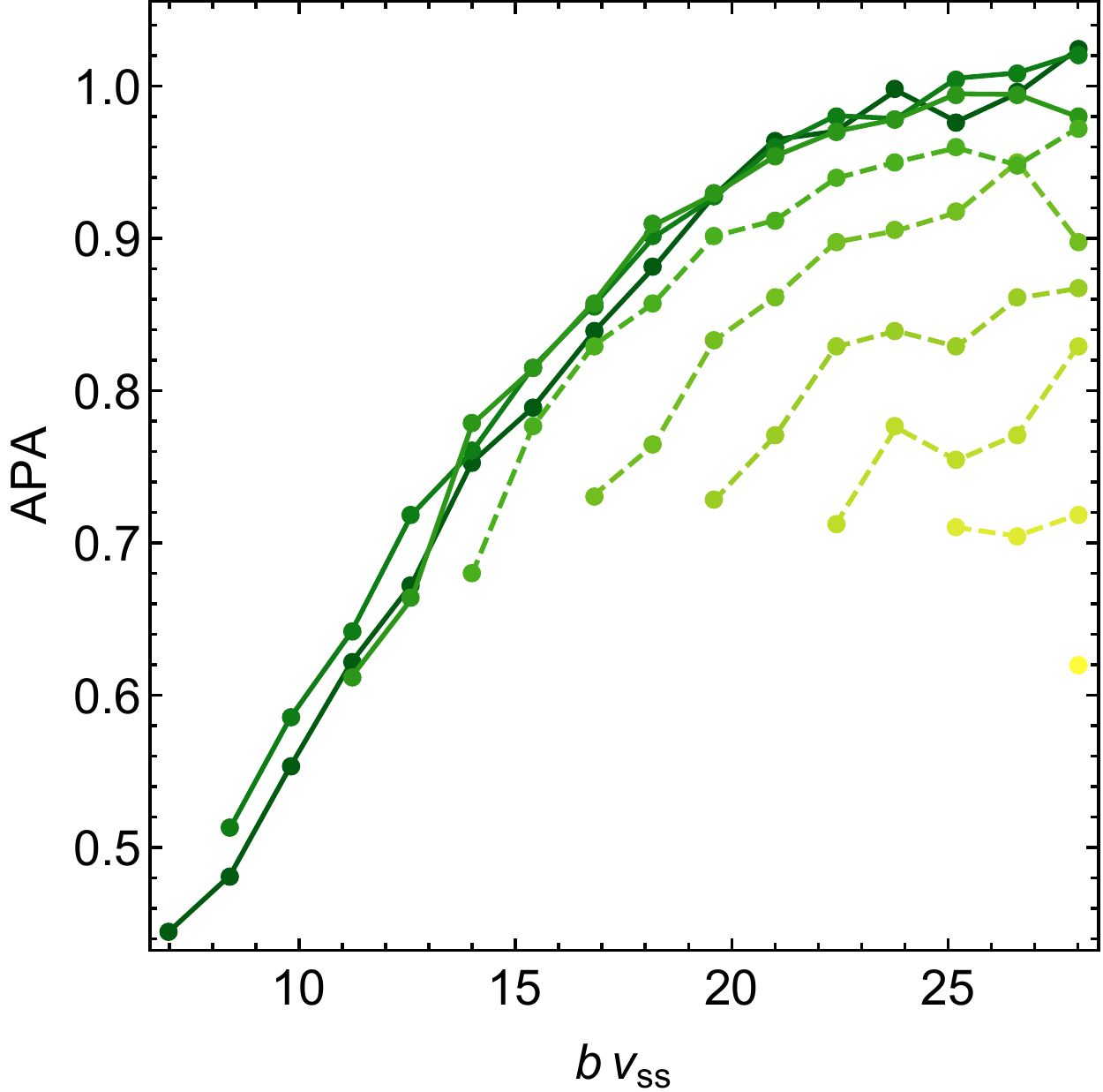}
  \caption{\footnotesize{Anterior-posterior asymmetries vs drag forces for a fixed target shape index of KV ($p_0^{KV}=3.8$) and various target shape index values of the surrounding cells (bottom to top: $p_0=3.2,3.3,3.4,3.5,3.6,3.7,3.8,3.9,4.0$). Solid lines correspond to the fluid-like tissues where $p_0>3.8$ while the dashed lines correspond to the solid-like tissue for values of $p_0 < 3.8$. Other simulation parameters are same as in Fig.~\ref{APA}.}}
  \label{APAvsDrag} 
 \end{figure}

To investigate the role of the drag forces on KV cell shape changes, we keep the mechanical properties (target shape index $p_0^{KV}$) of the KV cells fixed while varying the mechanical properties of the external cells and the speed of KV. The solid lines in Fig.~\ref{APAvsDrag} shows that APA collapses and is roughly proportional to the drag force in the fluid-like tissues where $p_0>3.8$, indicating the asymmetric cell shape changes within the KV can be driven by the drag forces. 

For values of $p_0 < 3.8$, indicated by dashed lines in Fig~\ref{APAvsDrag}, the tailbud tissue in our model is a soft solid with a yield stress. This means that if the propulsive force is sufficiently large, the KV can still push its way through the solid-like material.  However, it is clear that the data no longer collapse, as the force balance that generates the KV shape involves a complicated balance between the effective elasticity of the KV and that of the surrounding tissue.  

Although this is an interesting avenue for future study, our model suggest that this regime of parameter space with $p_0 < 3.8$ largely corresponds to APA values which are smaller than that seen in experiments.  Moreover, the ferrofluid droplet experiments in zebrafish tailbud suggest the tissue is fluid-like \cite{Serwane2016a}. Taken together, these results indicate that drag forces generated by surrounding fluid-like tailbud tissue could contribute to observed KV shape remodeling. 

 
\section*{Discussion}
We developed a vertex-like model to study the dynamical forces during development of the left-right organizer (KV) of the zebrafish embryo. By constraining the model parameters with our \textit{in vivo} measurements of the KV speed relative to surrounding tissue, as well as the rheological data of the zebrafish tailbud from Serwane et al. \cite{Serwane2016a}, we found that viscous flow of the cells around the KV generates a drag similar to a viscous fluid flowing past an elastic sphere, and the estimated magnitude of these forces is sufficient to drive cell shape changes observed in the KV.

More broadly, there are many morphogenetic processes in development that are apparently tightly controlled, but where the mechanisms that control cell shape remains unclear. Currently almost all of the proposed mechanisms and models involve shape changes that are either explicitly programmed into individual cells or dependent on time-invariant properties at the level of single cells, such as interfacial tensions. Very recent work emphasizes that large-scale, dynamic forces generated as groups of cells move through tissue can alter cell motility~\cite{Cai2016b,Aranjuez2016}. Here we take those ideas further by suggesting these forces can also also remodel cell shapes and control seemingly more complex morphogenetic processes.

These results highlight a challenge for modelers and scientists who study tissues and active matter. It is clear that the relaxation times measured in our droplet simulations, where deformations are small and induce no cellular rearrangements, are quite different from the relaxation times that are measured as we force the KV through the tissue. In the latter case, the motion of the KV results in finite deformations with large-scale cell motion and cellular rearrangements.  This suggest that the linear response of the tailbud might be quite different from its response to finite deformations, and that it is important to develop better constitutive models that develop coarse-grained equations for tissue motion. 

Another counterintuitive result of this work is that the relatively slow motion of the KV through the tailbud is sufficient to generate reasonably large drag forces. This is only possible because the effective viscosity of the tailbud tissue is high, with a characteristic timescale of a minute~\cite{Serwane2016a}. Since the speeds and tissue properties we identified here are quite common in embryogenesis, we conjecture that dynamic drag may be a mechanism that helps to control the morphology of other motile organs or migrating groups of cells in development and beyond. Possible candidates for this type of analysis include the zebrafish lateral line~\cite{Haas2006,Ma2009a}, where elongation may reduce the viscous drag, human MCF-7 mammary carcinoma cells that invade with a morphology similar to the lateral line in zebrafish~\cite{Friedl2009a}, mammary morphogenesis~\cite{Ewald2008}, mechanics of head formation~\cite{Varner2010}, hair follicle asymmetries~\cite{Devenport2008,Cetera2017}, and neural crest cell migration~\cite{Kuriyama2014,Teddy2004}.

Our simulation results emphasize that drag forces generated by the dynamics of the moving KV give rise to asymmetric cell shape changes regardless of the mechanism that drives the KV through the tissue. Experimentally, the mechanism remains unclear. We observed that the KV cells located posteriorly are protrusive and possibly migratory (Suppl. Movie 2), suggesting they may be pulling the KV through the surrounding tissue. Alternatively, this posteriorly directed movement could also be a result of convergence and extension, narrowing the embryo's mediolateral axis and lengthening its anterior-posterior axis. There could also be other mechanisms involved in the extension of the midline~\cite{Glickman2003}, or ECM anchoring the KV to the tailbud edge.

Another important question is whether a coupling between cell shape and cell motility could give rise to our observations, instead of tissue drag. Previous simulations using a similar model showed that cells tend to elongate slightly in the direction perpendicular to the direction of cell motility \cite{Giavazzi2018}. This effect is not dominant here, as the cell shape changes are not the same anteriorly (where cells are elongated parallel to the direction of motion) and posteriorly (where cells are elongated perpendicular to the direction of motion). These observations suggest that a simple coupling between cell motility and shape is not sufficient to generate the asymmetries we observe.

Our model assumes the tailbud tissue is confluent, with no gaps between cells. In the zebrafish tailbud, it is possible that there may be small gaps between cells, in which case the viscoelasticity will be controlled by the packing density in addition to cell shape~\cite{Teomy2018, Boromand2018}. Nevertheless, we expect the results developed here are still broadly applicable, as the drag forces generated by both confluent and non-confluent models should be fairly similar.

We have developed a two-dimensional model for KV motion through the tissue, but the KV and tailbud are manifestly three-dimensional tissues. In a three dimensional fluid at low Reynolds numbers ($Re$), the drag force on a moving sphere is given by Stokes's law: $D_{3D}=6 \pi \mu U L$, where $\mu$ is the viscosity, $U$ is the flow velocity of the sphere and $L$ is the radius of the sphere. In constrast, the drag force due to a 2D Stokes flow scales with the $Re$, $D_{2D} \sim \frac{4 \pi \mu_2 U}{-\ln(Re/4)}$~\cite{Lamb1932}.  $Re$ can be estimated by a simple dimensionless analysis as $Re=U \tau / L\approx0.1$ in natural simulation units with $U\approx0.2$, the viscous timescale $\tau=1$ as measured by the droplet deformation in our simulations and $L\approx2$ (radius of the KV in units of cell length). We can roughly estimate that the 2D viscosity is related to the 3D viscosity via the relevant length scale, $L \mu=\mu_2$. This suggests the drag force we estimate in 2D $D_{2D}$ may be about a factor of 5 lower than the relevant drag force in 3D, and therefore drag effects should be even more pronounced in 3D. Future work will extend a recently developed 3D model~\cite{Merkel2018} to the KV geometry. 

 The KV is symmetric at early stages of the development and advances to its final asymmetric structure over developmental time. This remodeling starts around 3-4 ss and becomes fully asymmetric after 8 ss~\cite{Wang2011}. Our model predicts that the asymmetry, quantified by APA, should track the drag forces, and therefore an open question is what might change the drag force as a function of time. One possibility is a change in speed of the KV relative to the surrounding tissue. Although there are no obvious differences in the speed of the KV relative to the yolk between 2 and 8 ss, it is possible that the retrograde flow of tailbud cells does change, and that will be a focus of future study. Alternatively, a global change in the material properties of the surrounding tissue would also change the drag force, and this should be investigated as well.

A related question is how long the KV remains asymmetric. In this study, we focus on relatively early stages of KV development (2 ss-8 ss) where the KV undergoes asymmetric cell shape changes. At slightly later stages (6ss-10ss), the cilia generates an asymmetric fluid flow inside the lumen which is required for its function, and at an even later stage (18ss) KV cells dissociate and move throughout the organism~\cite{Essner2005}. During the intermediate stage when fluid flow is required for downstream left-right patterning, the notochord and midline of the embryo are still extending, and so it is possible the KV remains asymmetric due to dynamic drag forces. It might be also possible that more stable ECM or mature adhesion complexes help to lock in the asymmetric structure after it forms dynamically. It would be interesting to analyze KV speed and cell shapes during the later stages of KV development in future studies. 

An important question is how this proposed mechanism might be tested more directly in experiments. One obvious candidate is drugs, knockdowns, or mutations that interfere with N-cadherin. Zebrafish tissue becomes less stiff as a result of impaired N-cadherin function~\cite{Serwane2016a}, and our model suggests a change in tissue stiffness will alter the asymmetries of the KV. However, N-cadherin mutations are also known to generate defective body axis extension~\cite{Lele2002,Harrington2007} and the defects in the extension might also impact the KV speed as the notochord is a midline structure. Thus, it may not be possible to distinguish between these effects.

Another obvious intervention would be to interfere directly with notochord extension, for example using a laser to ablate part of the notochord, or studying loss-of-function mutations which cause axis elongation defects. We hypothesize that these experiments would slow the motion of the KV through the tissue and cell shape changes might be disrupted. 

An important caveat to these proposals is that we know many important developmental processes are governed by partially redundant pathways, so that even if one pathway fails the organism can survive~\cite{Wagner2005}. For example, we and others have proposed that differential interfacial tensions generated via localization and activation of cytoskeletal molecules and ECM deposition~\cite{Wang2012,Compagnon2014a}, as well as differential changes in volume~\cite{Dasgupta2018}, can also drive cell shape changes. It is possible that these mechanisms work together with drag forces to robustly generate correct left-right patterning.


\section*{Acknowledgments}
We would like to thank Matthias Merkel, Daniel Sussman and Agnik Dasgupta for fruitful discussions. This work was supported by NIH grant R01GM117598 (G.E.T., J.D.A., M.L.M.). Additional support was provided by a grant from the Simons Foundation 446222 (G.E.T. and M.L.M.). Computing infrastructure support was provided through NSF ACI-1541396.

\bibliography{KVdynamics}
\setlength{\parskip}{1pt}
\beginsupplement
\section*{Supplementary Material}
\subsection*{Supplementary Methods}
\begin{figure}[ht]
 \centering
   \includegraphics[width=7in]{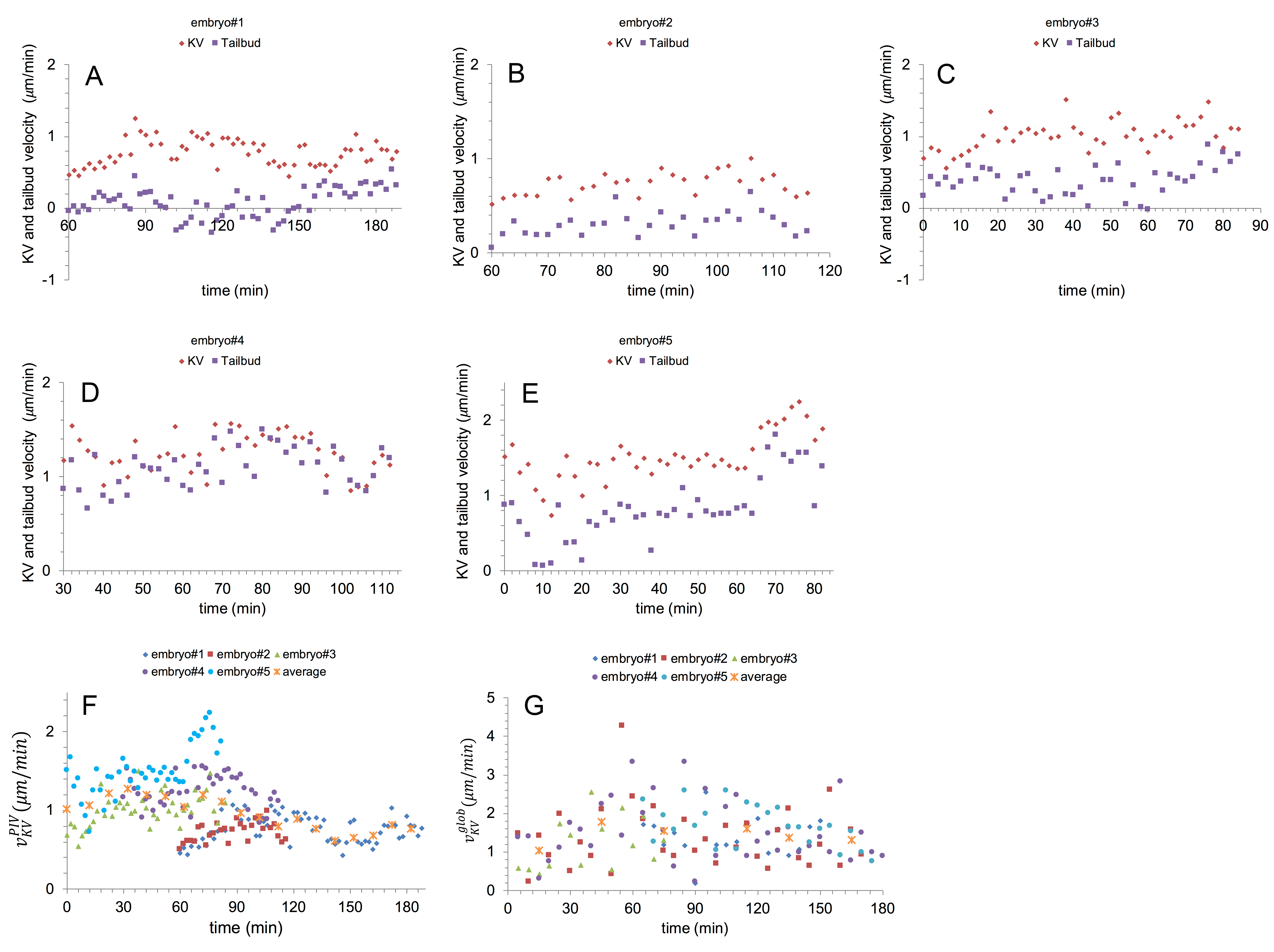}
   \caption{\footnotesize{Data for the velocities of KV and tailbud cells extracted from Particle Image Velocimetry (PIV) (A-F). (A-E) The mean component of the velocity along the posterior axis for points within the KV ROI (red dots) and tailbud ROI (purple squares) as a function of time for 5 different embryos.  The x-axis is defined so that t=0 corresponds to 2ss. (F) The absolute velocity along the posterior axis between the KV ROI and the tailbud ROI for each embryo, as well as the average value binned across 10 minute intervals(orange x's). (G) Absolute speed of the KV measured using the global imaging method as a function of time for 5 embryos. Orange x symbols in (G) represent the average of the data for every 30 minute intervals from all embryos. The 5 embryos in (A-F) are different than the 5 embryos in (G).}}
  \label{SuppFig-Speed} 
 \end{figure}

\begin{figure}[h]
\centering
   \includegraphics[width=2.8in]{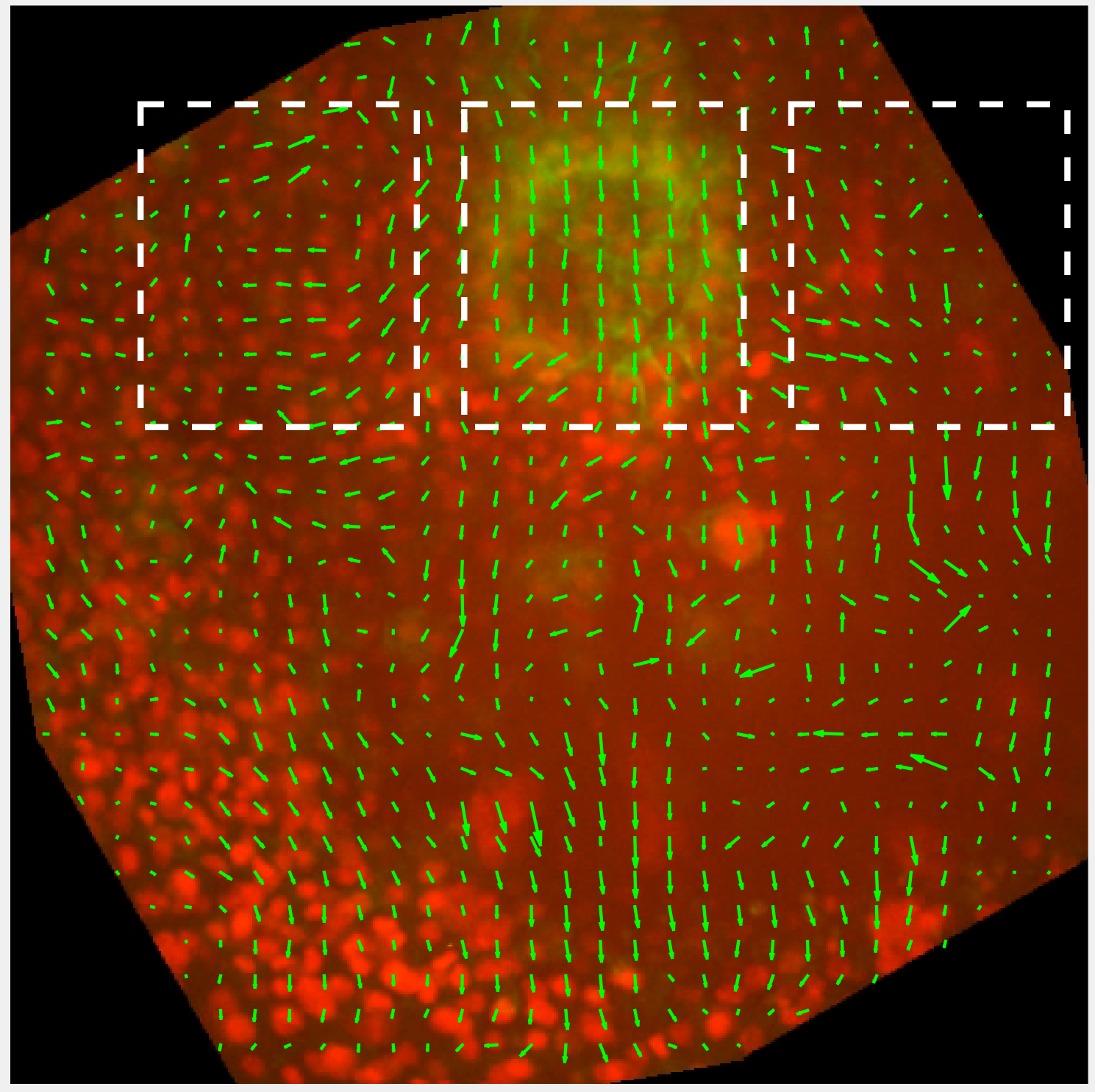}
   \caption{\footnotesize{A snapshot from Supplemantary Movie 2. The direction of cell flow is calculated using PIV. Green arrows represent the flow direction at every grid point. White dashed boxes are region of interest (ROI) where we calculate the mean velocity.}}
  \label{SupplMethod} 
 \end{figure}

We have measured the mean velocity of tissue regions including the KV and the surrounding tailbud tissue using Particle Image Velocimetry (PIV)~\cite{Thielicke2014,Thielicke2014a}. Specifically, we have defined three regions of interest (ROIs) (Fig.~\ref{SupplMethod}, white dashed boxes): one centered on the KV cells, and two centered on tailbud tissue on the right and left side of the KV. We have defined the ROI for the tailbud tissue as being shifted by about one cell diameter (10 $\mu m$) from the KV ROI. The width of the ROI for the KV is adjusted over time to account for lumen expansion, and the ROI for the surrounding tissue is fixed at 7 cell diameters wide. We use standard PIV software~\cite{Thielicke2014,Thielicke2014a} to identify the displacement at each material point between frames. Within each ROI, we calculate the mean velocity of the ROI as the vector average of the displacements of all grid points in the ROI divided by the time between frames. As the KV ROI moves almost entirely along the posterior axis, and we are interested in quantifying how quickly the KV moves relative to the surrounding tailbud tissue, we calculate and report only the component of these velocities along posterior axis.  Fig.~\ref{SuppFig-Speed}(A-E) is the mean velocity of the KV and the tailbud ROIs along the posterior axis as a function of time for 5 different embryos.  The relative velocity between the KV and tailbud ROIs along the posterior axis is shown in Fig. 2 (H) in the main text. By averaging this value over the entire time interval and all embryos, we find that the average relative velocity between KV and tailbud cells along the posterior axis is $v_{KVrel}^{PIV}= 0.55\pm 0.30 \;\mu m/min$. Fig.~\ref{SuppFig-Speed}(F) shows the absolute KV velocity along the posterior axis using the PIV method, and for comparison Fig.~\ref{SuppFig-Speed}(G) shows the absolute KV velocity measured using the global, lower-resolution method.

We have also repeated this analysis for the total velocity (e.g. not just the component along the posterior axis). In that case, we find that the relative velocity between the KV and tailbud is $v_{KVrel-total}^{PIV}= 0.57\pm 0.28 \;\mu m/min$, directed $37 \pm 47$ degrees to the left of the axis for the right-hand region and $1 \pm 72$ degrees to the left of posterior for the left-hand region.  This suggests that tailbud regions also have a small components of motion along the axis perpendicular to the KV, which is consistent with vortex-like displacement fields observed in the tailbud regions.  A more detailed quantification of the full displacement fields is an avenue for future research.

\subsection*{Modeling the indirect forces on KV development}
For the simulations presented in the main text, we have modeled a directed body force on the KV cells by applying a self-propulsion speed, $v_0^{KV}$. However, in reality such a directed motion does not exist. Rather, it rises naturally from other forces. In this section, we investigate whether our results depend sensitively on the assumption of a body force.

During embryonic development, mesodermal cells continuously rearrange. In convergent extension, cells converge narrowing the tissue with respect to the anteroposterior axis, and their subsequent intercalation lengthen the tissue along the midline. Because the intercalation along the mediolateral axis increases, the movement velocities of converging cells are not uniform \cite{Glickman2003}. The notochord, a midline structure posterior to the KV, interacts with the KV cells, and possibly pushes it towards the posterior body as a result of convergence extension movements. In addition, the cells next to the notochord also converge as the notochord extends along the midline. This extension might be the natural pushing force on KV cells.

To mimic the converging cells and to see if such motions could also generate shape changes in the KV, we assign a velocity as a function of position to the notochord cells and the group of cells next to the notochord (Fig.~\ref{APAcomparison} (A)). Specifically, we set $v_{0i}(x)= v_0*(x_i-x_{cm})$ where $x_i$ is the position of cell i, $x_{cm}$ is the center of mass of the tissue and the value of $v_0=1$. 

These converging cells do propel the KV towards the posterior as illustrated in the Supp. Movie 5. The cells further from the midline travel faster, and the cells along the midline extend as a consequence. We then compare the effect of the direct ($v_0^{KV}=0.4$) and indirect forces on the KV remodeling. For both cases, we calculate the instantaneous velocity of KV by performing a central average over the time intervals. Fig.~\ref{APAcomparison} (B) is a comparison of APA values as a function of instantaneous velocities in the case of direct and indirect forces on the KV, computed from the average of 100 separate simulation runs. The fact that the asymmetries are slightly higher in the case of indirect forces could be the result of different drag mechanism generated by the larger object or competing forces between the notochord and surrounding tailbud cells. Nevertheless, the resulting shape changes and the dependence of APA on velocity are similar, suggesting that any forces that propel the KV through the surrounding tissue will generate the observed cell shape changes.

\onecolumngrid

\twocolumngrid

\begin{figure}[h]
\centering
   \includegraphics[width=3in]{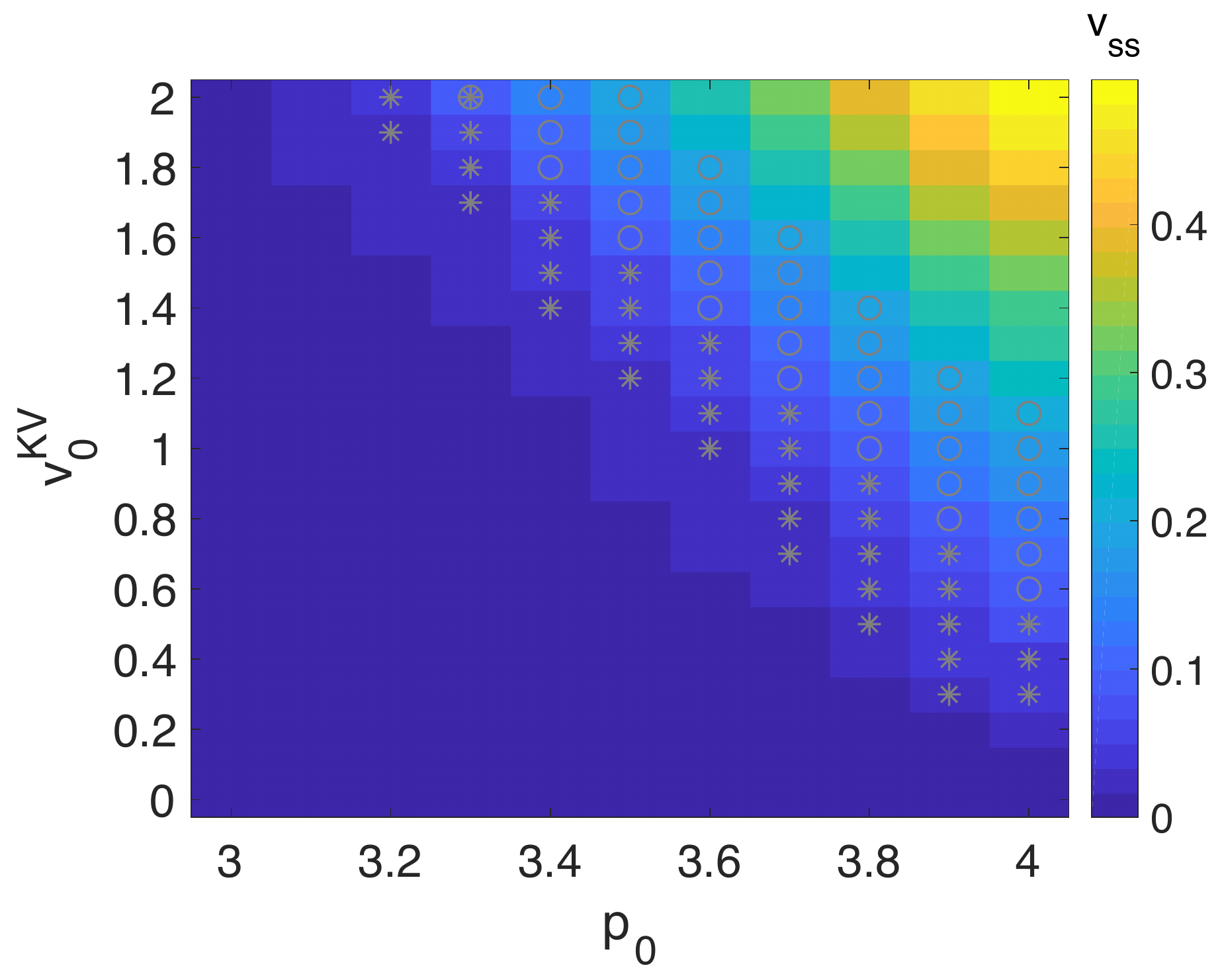}
  \caption{\footnotesize{Measured values of the steady-state speed of KV in the simulations as a function of self-propulsion speed of KV cells ($v_0^{KV}$) and target shape index of all cells ($p_0$). KV flows through the surrounding tissue in fluid-like environment whereas it stalls in the solid material. The region marked by asterisks corresponds to the average relative velocity of KV in the experiments in natural units ($v_{KVrel}^{PIV}= 0.055 \pm 0.030 \; l/\tau$) whereas the gray circles correspond to the global measurement of mean velocity in natural units ($v_{KV}^{glob}= 0.147\pm0.071 \; l/\tau$). Other simulation parameters are same as in Fig.~\ref{APA}.}}
  \label{vss} 
 \end{figure}
 
 \begin{figure}[h]
 \centering
\includegraphics[width=3.3in]{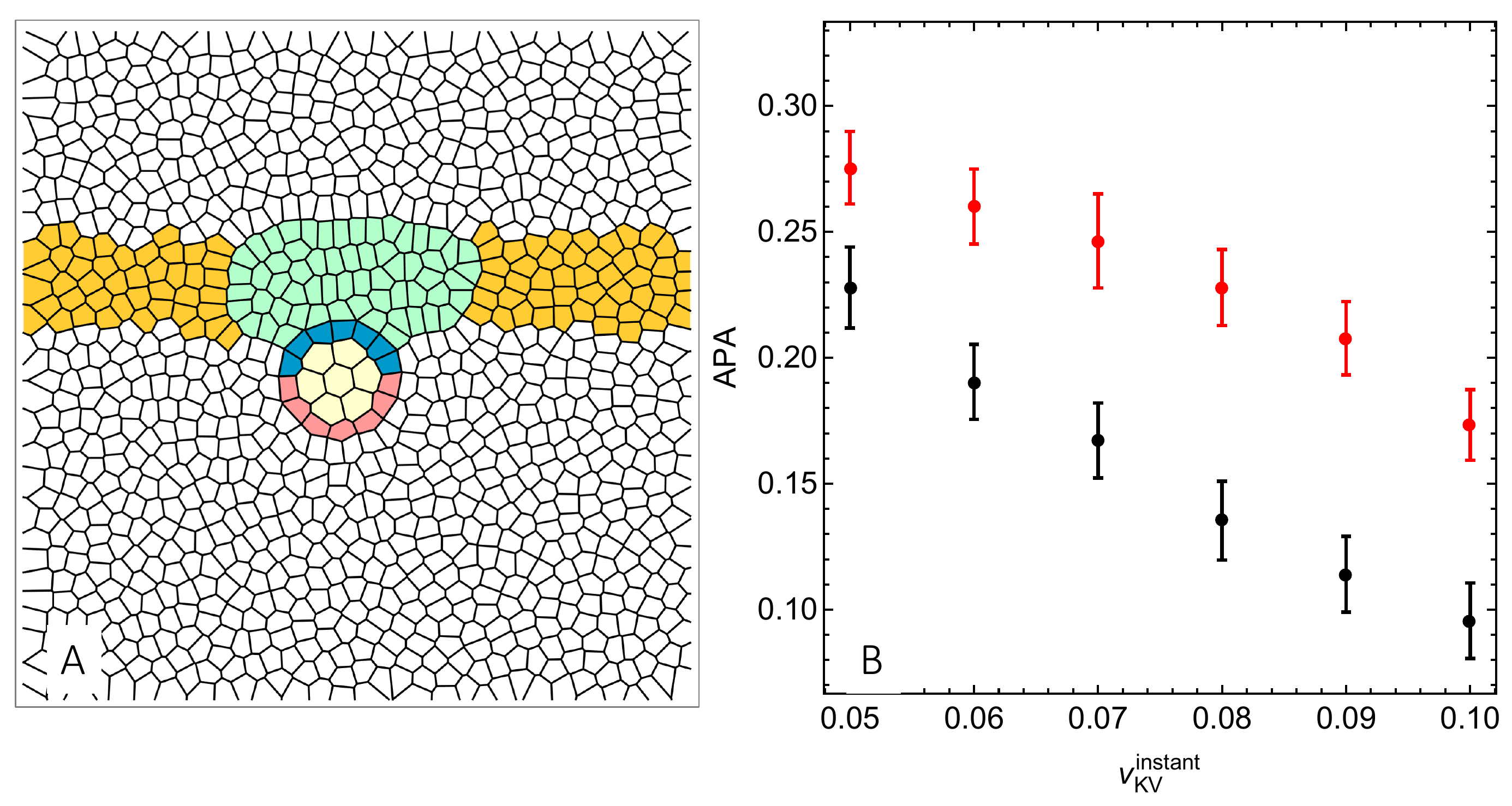}
\caption{\footnotesize{(A) Initial configuration of the model that assigns indirect forces on the KV. Green: notochord cells, yellow: cells next to the notochord, blue: anterior KV cells, pink: posterior KV cells, yellow: lumen Voronoi units, white: external cells. Simulation parameters are $p_0=3.8$, $A_0=1$, $n_{ext}=1000$, $n=7$, $n_{lumen}=7$, $K_A=100$, $K_P=1.0$, $\mu=1$, line tension between the KV cells and notochord cells $\gamma_{KV-NC}=2$, line tension between the KV cells and external cells $\gamma_{KV-ext}=2$, line tension between the notochord cells and external cells $\gamma_{NC-ext}=1$, line tension between the KV cells and lumen cells, $\gamma_{KV-lumen}=1$ and line tension between the external cells and lumen cells, $\gamma_{ext-lumen}=100$. (B) APA vs instantaneous velocities for the models that assign direct (black points) or indirect forces (red points) on KV for a fixed $p_0=3.8$. Data computed from the average of 100 separate simulation runs. Other simulation parameters for the model of the direct forces are same as in Fig.~\ref{APA}.}}
\label{APAcomparison} 
\end{figure}

 \begin{figure}[h]
 \centering
   \includegraphics[width=3.3in]{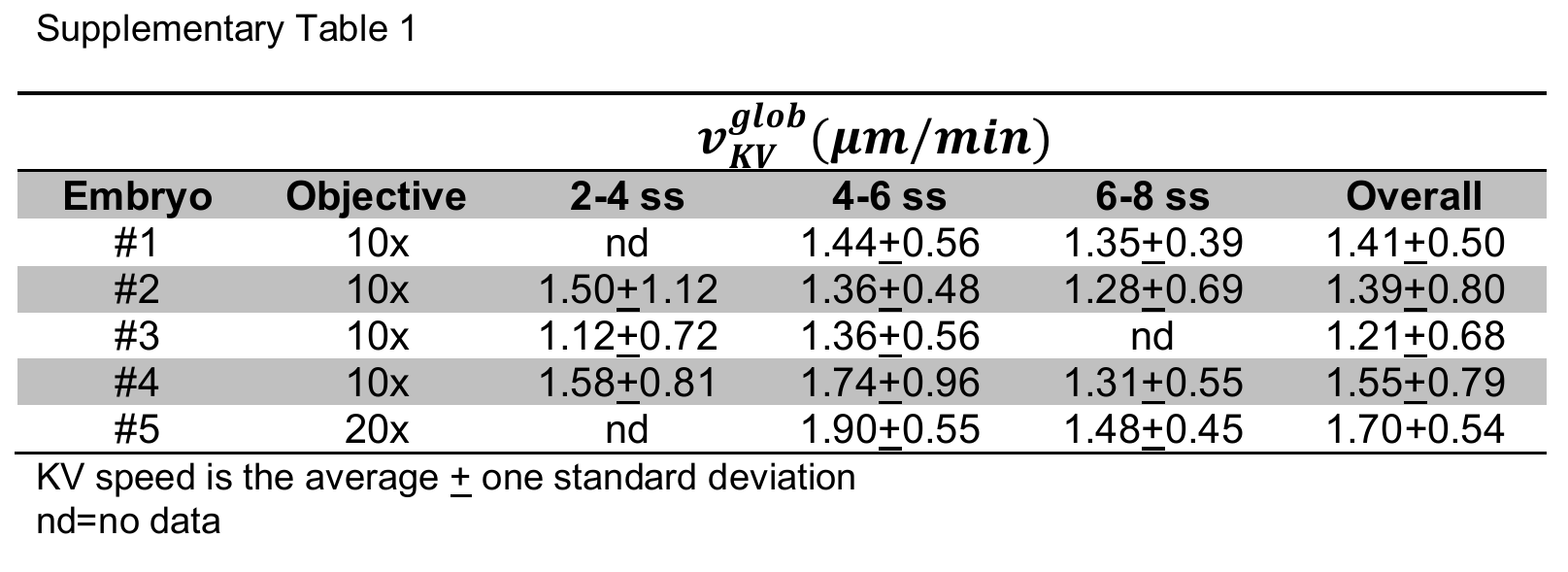}
  \label{SupplTable1} 
 \end{figure}
 
  \begin{figure}[h]
  \centering
   \includegraphics[width=3.3in]{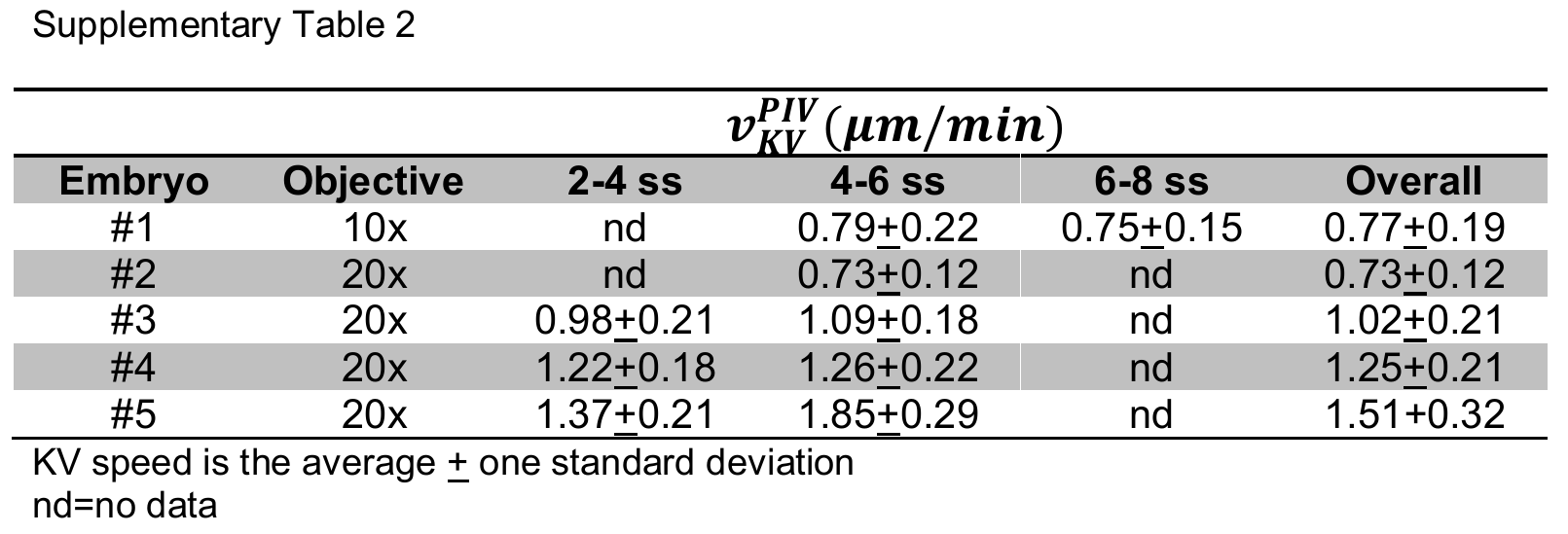}
  \label{SupplTable2} 
 \end{figure}
 
  \begin{figure}[h]
  \centering
   \includegraphics[width=3.3in]{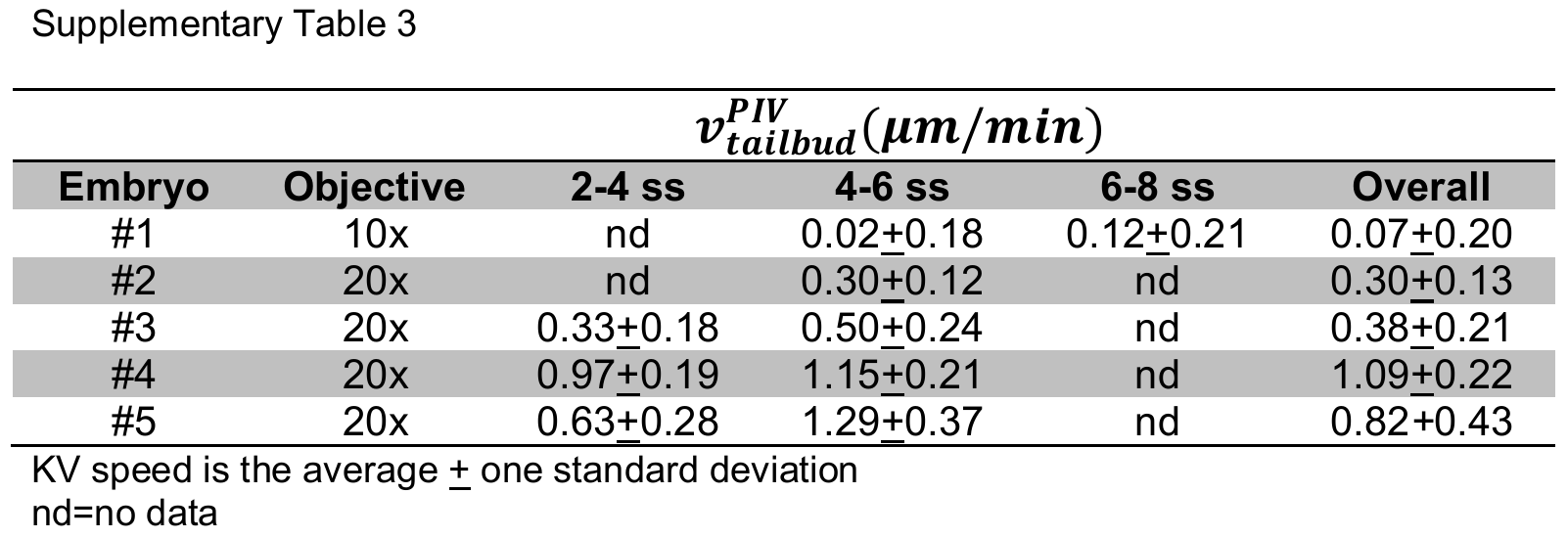}
  \label{SupplTable3} 
 \end{figure}
 
   \begin{figure}[h]
  \centering
   \includegraphics[width=3.3in]{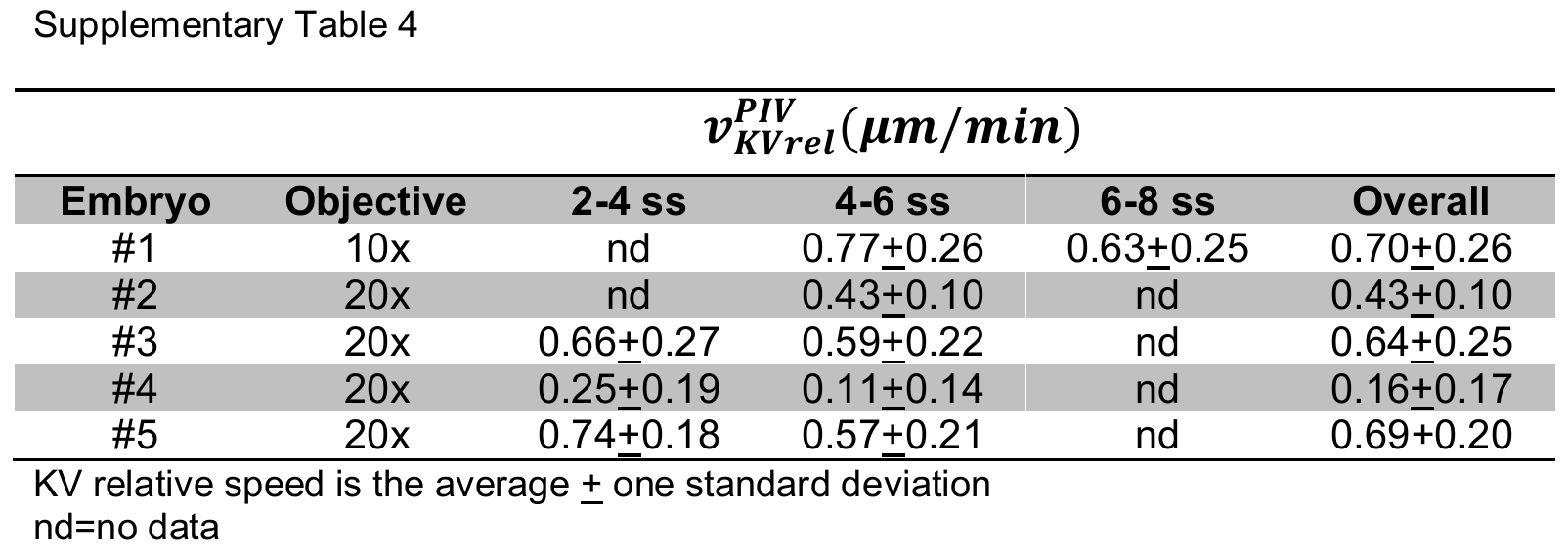}
  \label{SupplTable4} 
 \end{figure}

\clearpage
\section*{Movie Legends}
\small{
{\bf Supplementary Movie 1. Low-resolution time-lapse imaging of KV movement.} A representative set of experimental images used to track fluorescently labeled KV in live a transgenic Tg(sox17:GFP-CAAX) embryo between the 4 and 8 somite stages. Timestamp=hour:minute.

\noindent
{\bf Supplementary Movie 2.  High-resolution time-lapse imaging of KV movement.} A representative set of experimental images with all nuclei labeled with nuclear-localized mCherry, and KV cells labeled by green fluorescent protein in a wild-type Tg(dusp6:memGFP) embryo. Frames are taken at 2 minute intervals. Each frame is from a maximum projection and does not highlight KV cell shapes in the mid-plane as shown in Fig. 1(B).

\noindent
{\bf Supplementary Movie 3. Deforming a droplet embedded in a confluent tissue using Voronoi model.} Upon application of a constant force, the droplet relaxes to an ellipsoid along the direction of the force and likewise it goes back to its original spherical shape after the force is turned off. The force is applied over periodic cycles in 90$^\circ$ different directions.

\noindent
{\bf Supplementary Movie 4. KV is propelled through the surrounding tissue.} KV motion generates asymmetric cell shape changes that the anterior cells (blue cells) become elongated while the posterior cells (pink cells) become short and squat. Self-propulsion speed of the KV cells is $v_0^{KV}=1.4$ and target shape index of all cells is $p_0=3.8$. See Fig.~\ref{APA} for the other model parameters.

\noindent
{\bf Supplementary Movie 5. Modeling indirect forces on KV.} Converging cells propel the KV towards the surrounding tissue. See Fig.~\ref{APAcomparison} for the details of the topology and simulation parameters.}

\end{document}